\documentclass[lettersize,journal]{IEEEtran}
\usepackage{amsmath,amsfonts}
\usepackage{algorithmic}
\usepackage{array}
\usepackage[caption=false,font=normalsize,labelfont=sf,textfont=sf]{subfig}
\usepackage{textcomp}
\usepackage{stfloats}
\usepackage{url}
\usepackage{verbatim}
\usepackage{graphicx}
\usepackage{orcidlink}
\usepackage{multirow}
\usepackage{cite}
\def\BibTeX{{\rm B\kern-.05em{\sc i\kern-.025em b}\kern-.08em
    T\kern-.1667em\lower.7ex\hbox{E}\kern-.125emX}}
\usepackage{balance}
\begin{document}

\title{\fontsize{16}{20}\selectfont \textbf{EULER-ADAS:} \textbf{E}nergy-Efficient \& SIMD-\textbf{U}nified \textbf{L}ogarithmic-Posit \textbf{E}ngine\\ for Precision-\textbf{R}econfigurable Approximate ADAS Acceleration}

\author{Mukul Lokhande~\textsuperscript{\orcidlink{0009-0001-8903-5159}},~\IEEEmembership{Member, IEEE}, Ratko Pilipovi\'{c}~\textsuperscript{\orcidlink{0000-0002-4346-5487}}, Omkar Kokane\textsuperscript{\orcidlink{0009-0000-6288-7231}},\\
Adam Teman~\textsuperscript{\orcidlink{0000-0002-8233-4711}},~\IEEEmembership{Senior Member, IEEE},
and Santosh Kumar Vishvakarma~\textsuperscript{\orcidlink{0000-0003-4223-0077}},~\IEEEmembership{Senior Member, IEEE} 
        
\thanks{Manuscript received XXXX; revised XXXX.}
\thanks{
This research was supported by the Slovenian Research Agency under Grant P2-0359 (National Research Program on Pervasive computing), and Israeli Ministry Science, Innovation and Technology (MOST) under the Project wAIve. This work was also supported by the Special Manpower Development Program for Chip to Start-Up (SMDP-C2S), Ministry of Electronics and Information Technology (MeitY), Government of India, under Grant EE-9/2/21-R\&D-E.}
\thanks{Mukul Lokhande, Omkar Kokane, and Santosh K. Vishvakarma are with the NSDCS Research Group, Dept. of Electrical Engineering, Indian Institute of Technology Indore, Indore, India. Ratko Pilipovi\'{c} is with the Faculty of Computer and Information Science, University of Ljubljana, Ljubljana, Slovenia. Adam Teman is with EnICS Labs, Bar-Ilan Univ., Ramat Gan 5290002, Israel.}
\thanks{This work extends O. Kokane, M. Lokhande, G. Raut, A. Teman and S. K. Vishvakarma, "LPRE: Logarithmic Posit-enabled Reconfigurable edge-AI Engine", IEEE International Symposium on Circuits and Systems (ISCAS), London, United Kingdom, pp. 1-5, 2025. 
The prior conference publication focused on a reconfigurable logarithmic Posit arithmetic architecture that preserves application-level accuracy relative to FP32 while achieving significant hardware savings. This manuscript extends that work along three axes: combining logarithmic approximation with bit-truncation for accuracy-tunable mantissa arithmetic, adopting B-Posit representation to eliminate the variable-length encoding and decoding overhead, and introducing a SIMD-enabled accumulation datapath. To the best of our knowledge, this represents the first approximate logarithmic multiplier designed for the bounded Posit format. The extended evaluation further covers a comprehensive set of ADAS workloads to validate the proposed engine across representative application scenarios.
}

\thanks{Corr. author: Ratko Pilipovi\'{c} (e-mail: ratko.pilipovic@fri.uni-lj.si).}
}

\markboth{IEEE Transactions on Circuits and Systems for Artificial Intelligence,~Vol.~XX, No.~X, Month~202x}%
{Lokhande \MakeLowercase{\textit{et al.}}: Energy-Efficient \& SIMD-Unified  Logarithmic-Posit Engine for Precision-Reconfigurable Approximate ADAS Acceleration}

\maketitle

\begin{abstract}
Advanced driver-assistance systems (ADAS) require neural compute engines that deliver low-latency inference under strict power and area constraints. Posit arithmetic is attractive for such accelerators because it provides high numerical fidelity at low precision, but its variable-length regime encoding increases encode/decode cost and exposes the datapath to large regime-field fault effects. This paper presents EULER-ADAS, a SIMD-enabled logarithmic bounded-Posit neural compute engine for energy-efficient and reliability-aware ADAS acceleration. The proposed datapath combines bounded-regime Posit representation, stage-adaptive logarithmic mantissa multiplication with bit truncation, and a SIMD-shared quire accumulation path supporting Posit-(8,0), Posit-(16,1), and Posit-(32,2) execution. The unified architecture enables 4$\times$Posit-8, 2$\times$Posit-16, or 1$\times$Posit-32 operation without duplicating precision-specific hardware. FPGA implementation shows that the proposed configurations reduce LUT count by up to 41.4\%, delay by up to 76.1\%, and power by up to 71.9\% relative to exact Posit neural compute engines, while achieving up to $10\times$ lower energy-delay product than radix-4 Booth-based Posit multipliers. In 28-nm CMOS, the bounded variants occupy 0.013-0.016~mm\textsuperscript{2}, consume 19.8-22.1~mW, and operate at up to 1.84~GHz. Application-level evaluation across image-classification, ADAS, and edge-inference workloads shows that the evaluated Posit-16 and Posit-32 configurations remain within about 1.5 percentage points of FP32 accuracy. A Tiny-YOLOv3 prototype on Pynq-Z2 achieves 78~ms latency at 0.29~W and 22.6~mJ/frame, demonstrating the suitability of EULER-ADAS for low-power real-time ADAS inference.
\end{abstract}


\section{Introduction}

\begin{figure*}[!t]
    \centering
    \includegraphics[width=1.8\columnwidth]{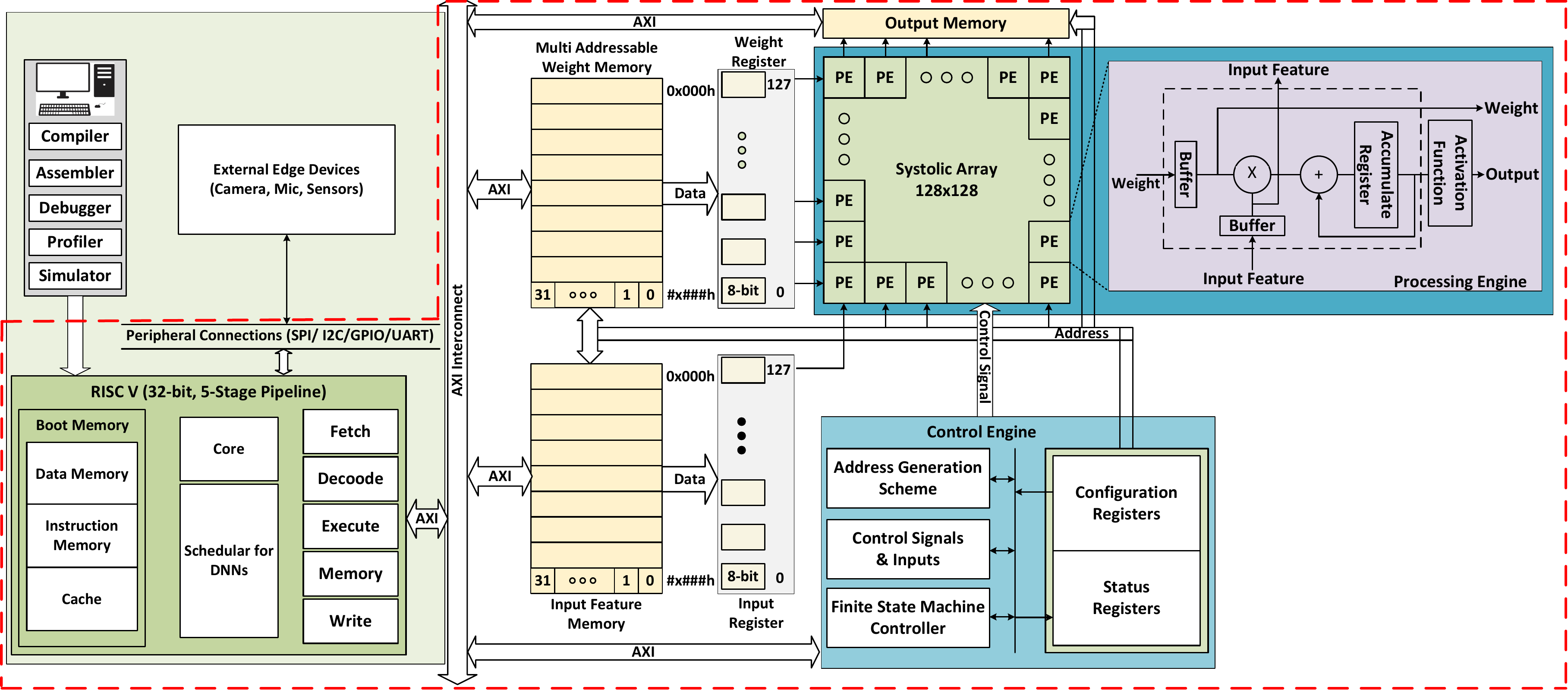}
    \caption{Domain-specific accelerator SoC architecture.}
    \label{fig:dsa_adas}
\end{figure*}

The deployment of autonomous vehicles and advanced driver-assistance systems (ADAS) relies on perception pipelines that must interpret complex driving environments under strict latency and power constraints. These pipelines increasingly depend on deep neural networks for tasks ranging from object detection~\cite{subin2023autonomous} to autonomous navigation~\cite{AD_TCAS}, motivating domain-specific accelerators (DSAs) that combine a host processor, on-chip memory hierarchy, and a dedicated neural compute engine (NCE), as illustrated in Fig.~\ref{fig:dsa_adas}. Within such accelerators, data movement and arithmetic dominate energy consumption: memory accesses can account for more than 80\% of system energy~\cite{TPUv4}, while multiply-accumulate (MAC) units occupy a large fraction of the arithmetic core area and active power~\cite{UVMAC}. Consequently, the numerical format used by the NCE becomes a central design lever, as it directly affects memory bandwidth, arithmetic complexity, and inference accuracy.

Low-precision quantization reduces memory traffic and MAC complexity by replacing IEEE~754 floating-point arithmetic with compact fixed-point formats, but this efficiency is often obtained at the cost of reduced numerical fidelity~\cite{MxP_TC, Rel1}. Posit arithmetic offers a more favorable accuracy-dynamic-range trade-off at low bit widths through variable-length regime field encoding (VLRE), enabling compact representations with dynamic range comparable to wider floating-point formats~\cite{Mallas2026Posit}. In DNN inference, Posit-8 and Posit-16 have been shown to match the accuracy of BFloat16 and FP32, respectively, while reducing memory footprint and improving arithmetic density~\cite{LPRE}. However, the same variable-length encoding that gives Posit its dynamic range also complicates hardware decoding and introduces reliability concerns, motivating a closer examination of Posit datapaths for ADAS accelerators.

For ADAS accelerators, numerical efficiency alone is insufficient: the arithmetic datapath must also remain robust under faults~\cite{11240678}. In standard Posit arithmetic, VLRE couples the numerical scale to a variable-length regime field, making the representation sensitive to bit flips in the regime bits. A single regime-field fault can therefore produce a disproportionately large numerical deviation, which is particularly undesirable in safety-relevant perception workloads. Fixed-regime Posit variants reduce this sensitivity~\cite{9399648}, but they do so by sacrificing much of the dynamic-range advantage that motivates Posit arithmetic in the first place.

Bounded Posit (B-Posit)~\cite{jonnalagadda2026closinggapfloatposit} provides a more balanced alternative by limiting the maximum regime length rather than removing regime variability entirely. This bounded-regime representation reduces the severity of catastrophic regime-field errors and has been shown to improve soft-error resilience by up to 47.2\% relative to standard Posit~\cite{mishra:hal-05569142}. At the same time, bounding the regime simplifies operand decoding and result encoding, reducing the hardware overhead of Posit arithmetic. B-Posit therefore addresses two central obstacles to deploying Posit NCEs in ADAS accelerators: reliability under faults and the cost of variable-length encode/decode logic.

While B-Posit reduces the cost and fault sensitivity of Posit encode/decode logic, the arithmetic core of the NCE remains dominated by mantissa multiplication and accumulation. Exact radix-4 Booth multipliers provide high numerical fidelity but impose substantial area, delay, and power overhead, especially in multi-precision SIMD datapaths. Approximate computing offers an attractive alternative because neural inference workloads can tolerate bounded arithmetic error~\cite{Leon2025Survey}. Among approximate multiplier families, logarithmic multipliers are particularly well suited to compact NCE design because they replace multiplication with addition in the logarithmic domain~\cite{Zhang_logfloat}. However, applying logarithmic multiplication only to the mantissa path leaves other Posit overheads intact, and existing SIMD designs often optimize multiplication without jointly addressing encode/decode and accumulation costs.

This paper presents EULER-ADAS, a SIMD-enabled logarithmic bounded-Posit NCE for energy-efficient and reliability-aware ADAS acceleration. EULER-ADAS jointly optimizes operand decoding, mantissa multiplication, quire accumulation, and result encoding within a unified precision-reconfigurable datapath. The architecture uses bounded Posit representation to reduce variable-length encode/decode overhead and mitigate regime-field fault sensitivity, a stage-adaptive logarithmic multiplier with bit truncation to provide accuracy-cost tuning in the mantissa path, and a SIMD-shared accumulation datapath that supports Posit-(8,0), Posit-(16,1), and Posit-(32,2) execution without duplicating precision-specific hardware.

The main contributions of this paper are as follows:

\begin{itemize}
    \item We propose EULER-ADAS, the first approximate B-Posit neural compute engine that integrates bounded-regime Posit encode/decode, stage-adaptive logarithmic mantissa multiplication, and SIMD-shared quire accumulation into a unified datapath.

    \item We introduce a configurable logarithmic Posit multiplication framework in which the multiplier stage count and truncation width act as accuracy-cost knobs across Posit-8, Posit-16, and Posit-32 operating modes.

    \item We provide FPGA and 28-nm ASIC evaluations showing that the proposed co-designed datapath reduces area, delay, power, and EDP relative to exact Posit and prior approximate NCE baselines.

    \item We validate the architecture across image classification, ADAS, and edge-inference workloads, including a Tiny-YOLOv3 Pynq-Z2 prototype, showing that the hardware savings are achieved with limited application-level accuracy loss.
\end{itemize}

The remainder of this paper is organized as follows. Section~\ref{sec:related} reviews related work. Section~\ref{sec:design} presents the theoretical analysis and architectural design of EULER-ADAS. Section~\ref{sec:eval} details the evaluation methodology and results. Section~\ref{sec:conclusion} concludes the paper and outlines directions for future work.

\section{Background and Motivation}
\label{sec:related}

\subsection{Related Work}

MAC operations dominate the arithmetic workload of DNN inference, and the multiplier is typically the most expensive component of the MAC datapath in energy, delay, and area~\cite{10729229, 10980492}. Exact Booth multipliers and full-precision accumulation paths provide high numerical fidelity, but they often exceed the precision required by neural inference workloads~\cite{10457067, 10577667}. Since DNNs can tolerate bounded arithmetic perturbations, approximate multiplication has emerged as an effective approach for reducing hardware cost while preserving application-level accuracy~\cite{11363033}.

Approximate multipliers can be broadly grouped into non-logarithmic and logarithmic designs. Non-logarithmic approaches reduce the complexity of conventional multiplication by simplifying Booth encoding~\cite{Jiang2025EfficientApprox, 9997088, RAD1024_TVLSI}, using approximate compressors in the partial-product reduction tree~\cite{9920015}, decomposing operands into smaller segmented multipliers~\cite{10577667}, or representing values with stochastic bitstreams~\cite{10970099}. These techniques can reduce area and power, but their benefits are often limited by the remaining partial-product generation and reduction logic. Booth simplification is usually constrained to modest radix choices, compressor-based designs require careful error compensation or bit-position selection, segmented multipliers mainly approximate lower-significance product bits, and stochastic designs suffer from latency and scalability limitations as operand width increases.

Logarithmic approximate multipliers address some of these limitations by following Mitchell's product approximation~\cite{5219391}, which transforms multiplication into addition in the logarithmic domain using a binary logarithm approximation. This structure avoids explicit partial-product generation and can provide larger area and energy savings than many non-logarithmic approximations~\cite{ALM_SOA, LUT-ALM_TC, Ratko_TL, DRALM_TSC'22}. The main drawback is higher arithmetic error, which has motivated correction and refinement techniques~\cite{BABIC201123}. However, prior studies show that DNN inference can often tolerate the uncorrected error of logarithmic multipliers while maintaining accuracy comparable to exact or non-logarithmic approximate designs~\cite{MITCH_TRUNC_TC, Ratko_TL}.

The compactness of logarithmic multipliers has led to their use in floating-point~\cite{Zhang_logfloat, 10433081} and Posit arithmetic~\cite{PFP_ILM_approx}, where they primarily target mantissa multiplication. In Posit NCEs, however, multiplier-level savings can be limited by the surrounding datapath, including variable-length operand decoding, exponent and regime processing, result encoding, and accumulation. Resource-shared SIMD engines~\cite{C-SIMD, RPE} improve multi-precision utilization, but they do not jointly address Posit reliability, bounded-regime encode/decode complexity, approximate mantissa multiplication, and mode-shared accumulation. EULER-ADAS targets this combined design problem by co-designing a stage-adaptive logarithmic mantissa multiplier with a bounded-Posit datapath and a SIMD-shared quire accumulator.

\subsection{Design Framework for Reliability and Approximation}
\label{sec:framework}

This subsection establishes the analytical framework used to select the bounded-regime and logarithmic-multiplier parameters of EULER-ADAS. The framework links three design variables: the bounded regime length $R$, which controls the severity of regime-field faults; the logarithmic multiplier stage count $n$, which controls approximation error; and the truncation width $m$, which provides an additional area-accuracy trade-off.

\subsubsection{Reliability of Bounded Posit Representation}

A posit value is represented as
\begin{equation}
x = (-1)^s \cdot (\mathrm{useed})^k \cdot 2^e \cdot (1+f),
\qquad
\mathrm{useed}=2^{2^{e_s}},
\label{eq:posit_value}
\end{equation}
where $s$ is the sign bit, $k$ is the regime value, $e$ is the exponent, and $f$ is the fraction. Since the regime term dominates the dynamic range of posit numbers, any corruption in the regime field can produce a catastrophic change in numerical magnitude. Therefore, bounding the regime directly reduces the sensitivity of the representation to bit-flips. A bounded-regime posit is denoted as
\begin{equation}
\mathrm{bPosit}(N,e_s,R),
\end{equation}
where $N$ is the word length, $e_s$ is the exponent size, and $R$ is the maximum regime-field width in bits.

Following the soft-error resilience analysis for Posit arithmetic in~\cite{mishra:hal-05569142}, we characterize the fault sensitivity of bounded Posit using the Expected Catastrophic Error (ECE). ECE measures the expected distortion in logarithmic magnitude between the original value $x_o$ and the faulty value $x_f$:
\begin{equation}
\eta_B = \mathbb{E}\left[\left|\log_2|x_o| - \log_2|x_f|\right|\right],
\label{eq:ece_definition}
\end{equation}
The metric can be decomposed into regime and exponent contributions as
\begin{equation}
\eta_B \approx 2^{e_s}\,\mathbb{E}[|k_o-k_f|] + \mathbb{E}[|e_o-e_f|].
\label{eq:ece_decomposition_}
\end{equation}
The closed-form ECE for a bounded Posit configuration $(N, e_s, R)$ is given by:
\begin{equation}
\eta_B(N,e_s,R) = 2^{e_s}\left[G_1(R)+G_2(R)+G_3(R)\right] + \frac{2^{e_s}-1}{2},
\label{eq:ece_bounded}
\end{equation}
where $G_1(R)$, $G_2(R)$, and $G_3(R)$ denote the expected contributions from faults in the regime run bits, the regime terminating bit, and the exponent field, respectively~\cite{mishra:hal-05569142}. Since $\eta_B$ increases monotonically with the regime bound $R$, reducing $R$ lowers the expected catastrophic error:
\begin{equation}
R_1 < R_2 \implies \eta_B(N,e_s,R_1) < \eta_B(N,e_s,R_2).
\label{eq:ece_monotonic}
\end{equation}
The resilience improvement of bounded Posit relative to standard Posit is expressed through the improvement factor:
\begin{equation}
\Gamma_B(N,e_s,R) = \frac{\eta_{\mathrm{std}}(N,e_s)}{\eta_B(N,e_s,R)},
\label{eq:gamma_bounded_v1}
\end{equation}
where $\Gamma_B > 1$ indicates lower expected catastrophic error than standard Posit. The reliability benefit of bounding the regime comes primarily from limiting the numerical severity of regime-field faults, rather than from reducing the frequency of exceptional values. Thus, B-Posit improves resilience by constraining the magnitude distortion induced by a fault while retaining much of the dynamic-range flexibility of standard Posit.
A more detailed analysis of reliability of B-Posit representation is provided in the Supplemental Materials, which accompany paper.
\subsubsection{Approximation Error of Logarithmic Multiplication}

For the stage-adaptive logarithmic multiplier, the relative approximation error of an $n$-stage implementation is bounded by:~\cite{BABIC201123}
\begin{equation}
RE(n) \leq 2^{-2n},
\label{eq:ilm_error}
\end{equation}
with the worst case occurring when both multiplicands have all-one fractional patterns. When operand truncation is also applied, only $m$ bits after the leading-one position are retained. The combined relative error can then be bounded as
\begin{equation}
RE(n,m) \leq 2^{-2n} + 2^{-m},
\label{eq:combined_error}
\end{equation}
where the first term captures logarithmic approximation and the second term captures truncation error.

\subsubsection{Design-Space Exploration}

Equations~(\ref{eq:ece_bounded})-(\ref{eq:combined_error}) define the three design parameters used in EULER-ADAS: stage count $n$, truncation width $m$, and bounded regime length $R$. The stage count is selected to keep the logarithmic approximation error commensurate with the target precision: for 8-bit Posit, $n \in \{2, 3\}$ yields $RE \leq 2^{-4}$ and $2^{-6}$, respectively; for 16-bit Posit, $n \in \{4, 6\}$ provides tighter bounds for the wider mantissa field; and for 32-bit Posit, $n \in \{8, 12\}$ further reduces approximation error for high-precision execution. The truncation width is selected so that the truncation term $2^{-m}$ remains controlled relative to the logarithmic approximation term $2^{-2n}$, retaining 4 or 5 bits for 8-bit Posit, 8 or 10 bits for 16-bit Posit, and 16 or 20 bits for 32-bit Posit. Finally, the regime bound $R$ is chosen as the smallest value that preserves the dynamic range required by the target workloads, with $R = 2$ for Posit-8, $R = 3$ for Posit-16, and $R = 5$ for Posit-32. This choice improves fault resilience according to~(\ref{eq:ece_monotonic}) while retaining sufficient dynamic range for inference.

\begin{figure*}
    \centering
    \includegraphics[width=1\linewidth]{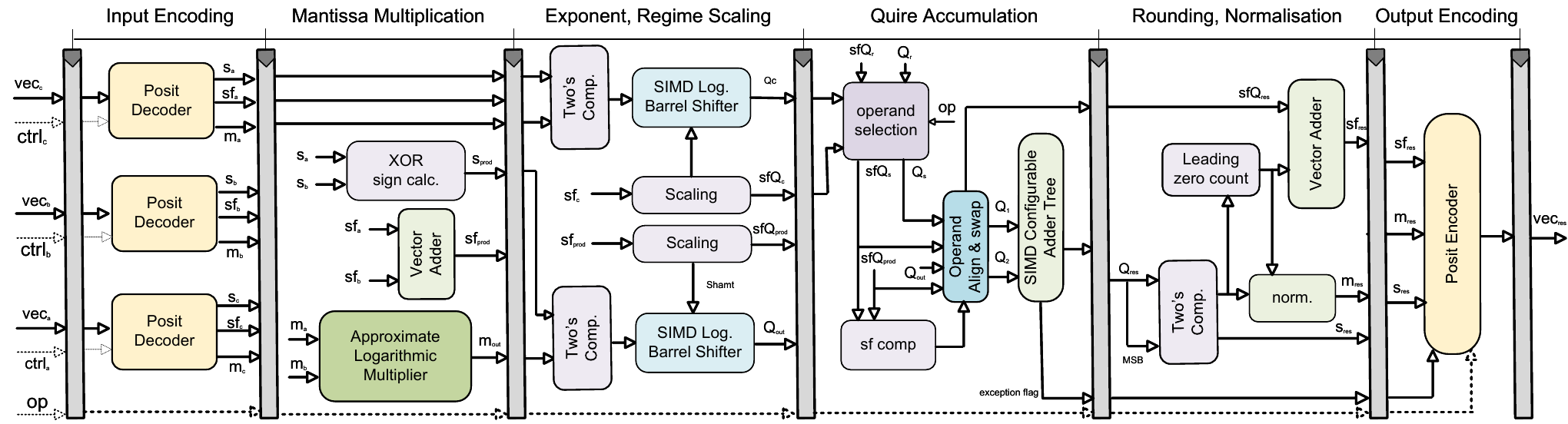}
    \caption{Detailed datapath of the unified SIMD Logarithmic-Posit neural compute engine (NCE), highlighting the key focus of this work: mantissa multiplication, the SIMD-configurable adder tree, and the decoder/encoder unit.}
    \label{fig:prop_NCE}
\end{figure*}

\section{EULER-ADAS: Architecture}
\label{sec:design}

The EULER-ADAS datapath, shown in Fig.~\ref{fig:prop_NCE}, implements a six-stage SIMD pipeline for precision-reconfigurable Posit multiply-accumulate execution. The datapath supports Posit-(8,0), Posit-(16,1), and Posit-(32,2) modes within a shared hardware structure, enabling multi-precision operation without duplicating precision-specific MAC units. Guided by the framework in Section~\ref{sec:framework}, the architecture combines bounded-Posit encode/decode logic, stage-adaptive logarithmic mantissa multiplication, SIMD exponent and regime scaling, quire-based accumulation, and final rounding and normalization. The six pipeline stages are described below.

\begin{figure*}
    \centering
    \includegraphics[width=1\linewidth]{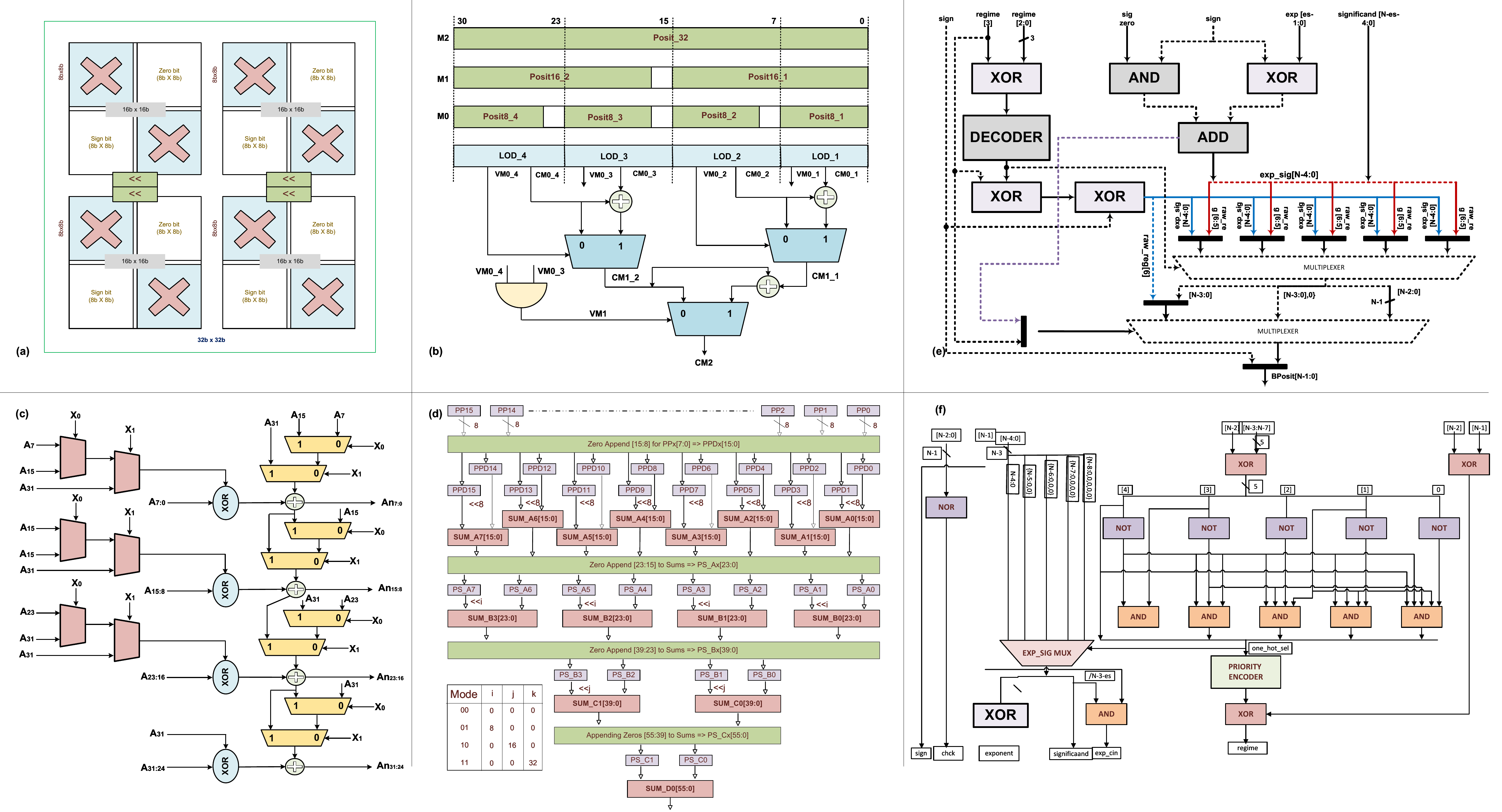}
    \caption{Detailed internal circuitry of SIMD-configurable compute blocks, illustrating: (a) high-precision split mantissa multiplication, (b) two's-complement circuitry, (c) leading zero/one detection, (d) restructured SIMD-configurable adder tree, (e) B-Posit encoder, and (f) B-Posit decoder.}
    \label{fig:IntCircs}
\end{figure*}

\vspace{0.35em}
\noindent\textbf{Stage 1: Operand Decoding.}
The SIMD input vectors $(\mathbf{vec\_a}, \mathbf{vec\_b}, \mathbf{vec\_c})$ are first decomposed into sign, regime, exponent, and mantissa fields using resource-shared bounded-Posit decoders~\cite{jonnalagadda2026closinggapfloatposit}. The decoding circuitry, shown in Fig.~\ref{fig:IntCircs}(f), exploits the bounded regime length to avoid the long variable-length extraction path of conventional Posit decoders. The regime bits are XORed with the sign bit and mapped into a one-hot representation using simple NOT and AND logic. This one-hot value selects the exponent scaling through the \textit{EXP\_SIG MUX}, while the regime count is recovered through a priority encoder.

Because bounded-Posit operands are stored in two's-complement form, sign-aware extraction is used to recover the constituent fields before arithmetic processing. The raw exponent and regime bits are XORed with the sign bit to generate the final exponent and regime values. As a result, the decoder has a largely bit-width-invariant combinational path, in contrast to standard Posit decoders whose critical paths include chained leading-bit detection, shifting, multiplexing, and addition~\cite{UVMAC}. This bounded-regime decoding strategy reduces the latency and hardware cost of the operand-processing stage, at the expense of a narrower dynamic range.

\begin{figure}[!t]
    \centering
    \includegraphics[width=\columnwidth]{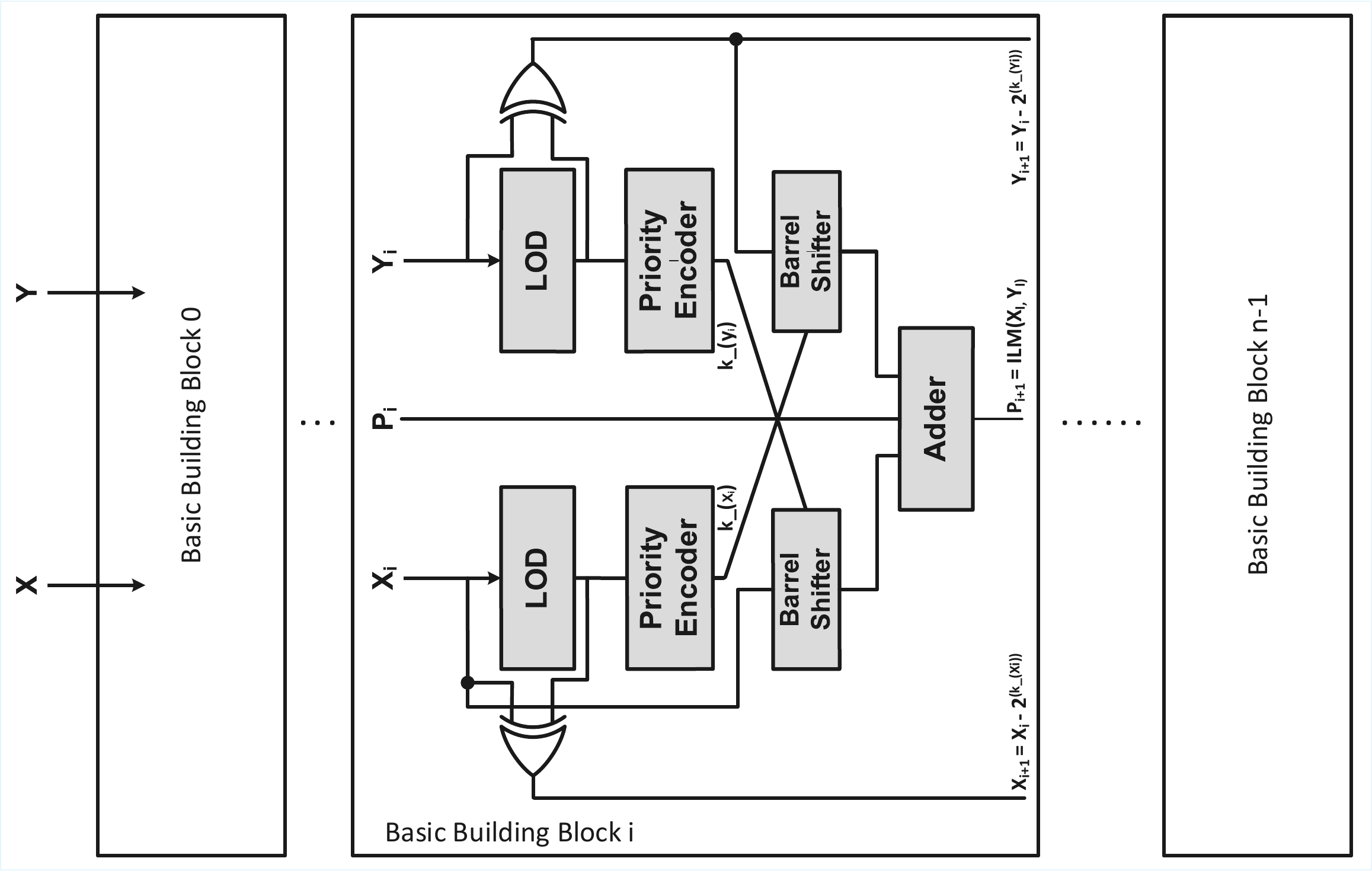}
    \caption{Pareto-optimal parallel stages for the stage-adaptive logarithmic multiplication approach.}
    \label{fig:parallel_ILM}
\end{figure}

\vspace{0.35em}
\noindent\textbf{Stage 2: Mantissa Multiplication.}
The decoded mantissa fields are processed by SIMD-configurable stage-adaptive logarithmic multipliers (ILMs)~\cite{BABIC201123}, as shown in Fig.~\ref{fig:parallel_ILM}. These multipliers replace the radix-4 Booth mantissa multipliers used in exact Posit datapaths~\cite{UVMAC}. Instead of generating and reducing partial products, the ILM approximates multiplication through logarithmic-domain addition, substantially reducing multiplier complexity. To further reduce area and power, operand truncation is applied after leading-one detection, retaining only the $m$ most significant bits required by the selected accuracy mode.

The SIMD multiplier organization follows the high-precision split strategy in Fig.~\ref{fig:IntCircs}(a), allowing the same datapath to support Posit-8, Posit-16, and Posit-32 execution without the operand-bandwidth limitations of low-precision combination schemes~\cite{RPE}. The stage count $n$ and retained width $m$ are selected according to the error bounds in Section~\ref{sec:framework}: Posit-8 uses 2 or 3 ILM stages with 4 or 5 retained bits, Posit-16 uses 4 or 6 stages with 8 or 10 retained bits, and Posit-32 uses 8 or 12 stages with 16 or 20 retained bits. The resulting Posit-level error metrics are reported in Table~\ref{tab:lpos_error}.

\vspace{0.35em}
\noindent\textbf{Stage 3: Exponent and Regime Scaling.}
The product sign is computed through parallel bitwise XOR logic, while the decoded exponent and regime fields are combined using the two-stage addition structure in~\cite{UVMAC}. The resulting scale value determines the alignment shift required before quire accumulation. To support all precision modes within the same datapath, EULER-ADAS uses a SIMD logarithmic barrel shifter~\cite{RFPU} that applies the required scaling across Posit-8, Posit-16, and Posit-32 lanes. The computed scaling factors are then forwarded to the accumulation stage.

\vspace{0.35em}
\noindent\textbf{Stage 4: Quire Accumulation.}
The accumulation stage uses the restructured SIMD-configurable adder tree shown in Fig.~\ref{fig:IntCircs}(d). Partial products from 8-, 16-, and 32-bit modes are accumulated into a shared 128-bit quire, with sign extension and two's-complement conversion applied according to the product sign and decoded regime information. The exponent and regime scaling factors determine the alignment shift before accumulation, allowing all supported precision modes to reuse the same reduction topology.

Because mode reconfiguration changes only lane partitioning, operand alignment, and shift distance, the accumulation delay is governed primarily by the adder-tree depth rather than by the number of supported precision modes. This avoids duplicating precision-specific accumulators while preserving sufficient quire width for the worst-case dynamic range across the supported Posit formats. The shared 128-bit quire also delays final rounding until after accumulation, reducing cumulative rounding error.

\vspace{0.35em}
\noindent\textbf{Stage 5: Rounding and Normalization.}
After quire accumulation, the result is normalized before being packed into the target Posit format. A SIMD-enabled leading-zero-count (LZC) module, shown in Fig.~\ref{fig:IntCircs}(c), determines the normalization shift, while the two's-complement circuitry in Fig.~\ref{fig:IntCircs}(b) restores the correct signed representation when required. Round-to-nearest-even is then applied using guard, round, and sticky bits to reduce quantization error at the selected output precision.

The control path handles zero, sign inversion, overflow, and special-value propagation, including flush-to-zero, denormals-are-zero, and $\pm$Inf/NaN cases according to the Posit-2022 handling used in the baseline design. Directed corner-case tests confirm that approximation is confined to mantissa multiplication, while normalization, rounding, and exception handling remain exact.

\vspace{0.35em}
\noindent\textbf{Stage 6: Result Encoding.}
In the final stage, the bounded-Posit encoder packs the normalized sign, regime, exponent, and mantissa fields into the output Posit word~\cite{LPRE}. As shown in Fig.~\ref{fig:IntCircs}(e), the bounded regime length allows the candidate regime, exponent, and fraction fields to be prepared in parallel and selected through a compact multiplexer structure. The regime width is determined by XORing the lower regime bits with the regime MSB, after which the intermediate regime string is decoded and corrected using the regime sign.

Compared with conventional Posit encoders, which rely on sequential control logic, adders, shifters, and binary decoders, the bounded-Posit encoder has a shorter and more regular critical path. In the implemented datapath, the critical path consists primarily of XOR logic, a binary decoder, and two multiplexer stages, reducing the output-packing overhead that would otherwise limit the benefit of approximate mantissa multiplication.

\section{Methodology \& Performance Evaluation}
\label{sec:eval}

\begin{figure*}[!t]
    \centering
    \includegraphics[width=\textwidth]{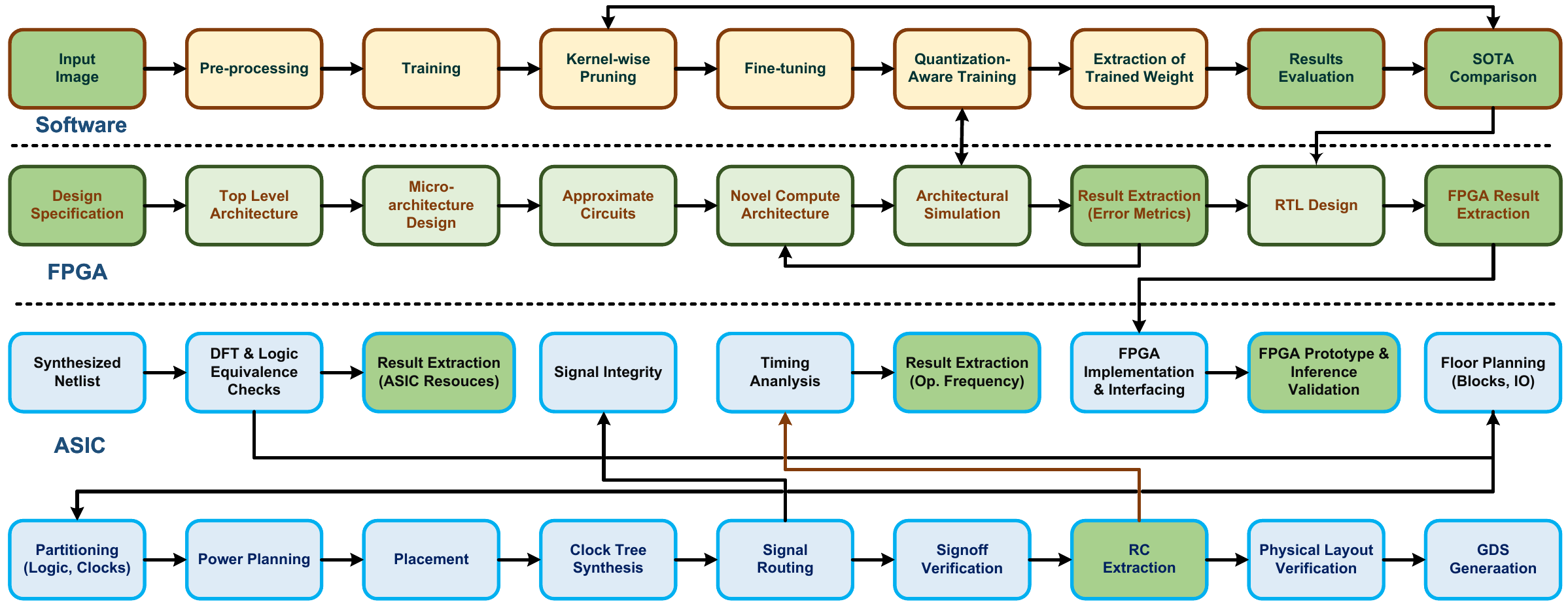}
    \caption{Detailed methodology/workflow illustrating algorithm-hardware co-design.}
    \label{fig:methodology}
\end{figure*}

The evaluation methodology, summarized in Fig.~\ref{fig:methodology}, combines arithmetic-level modeling, RTL validation, FPGA implementation, ASIC synthesis, and application-level inference experiments. At the arithmetic level, a Python bit-accurate model implements the proposed bounded-Posit and logarithmic-multiplier variants and is used to measure numerical error across scalar and SIMD configurations. The RTL implementation is written in Verilog HDL and verified in QuestaSim against the Python model using matched input vectors. FPGA results are obtained after implementation in AMD Vivado on Xilinx VC707 and PYNQ-Z2 platforms, while ASIC results are obtained using a Cadence RTL-to-GDS flow with Genus synthesis and Innovus place-and-route in a 28-nm CMOS HPC process. Application-level experiments use the same bit-accurate arithmetic model to evaluate the accuracy impact of the proposed configurations on representative ADAS and edge-inference workloads.

\subsection{Arithmetic Error Evaluation}

To quantify the arithmetic error introduced by the proposed approximations, four metrics are evaluated across all configurations: mean squared error (MSE), mean absolute error (MAE), normalized mean error distance (NMED), and mean relative error distance (MRED). Each configuration is denoted \texttt{b$x$\_LP-$n$\_T$m$}, where $x$ is the upper bound on regime bits, $n$ is the logarithmic multiplier stage count, and $m$ is the number of retained bits after truncation. Error metrics are obtained for logarithmic Posit multipliers in 2- and 3-stage configurations for 8-bit Posit, 4- and 6-stage configurations for 16-bit Posit, and 8- and 12-stage configurations for 32-bit Posit. Each variant is compared against exact radix-4 Booth-based Posit multiplication, and SIMD configurations are additionally evaluated in 8b/16b and 8b/16b/32b operating modes.

Table~\ref{tab:lpos_error} shows that increasing the logarithmic multiplier stage count generally improves arithmetic accuracy, while truncation introduces a modest additional error that can be traded for lower hardware cost. The lowest-error entries in each precision group are highlighted in bold. Increasing the retained width $m$ reduces truncation error, although its effect is smaller than the change caused by the logarithmic approximation stage count. Bounded-regime variants incur additional error because restricting the regime length narrows the representable dynamic range relative to standard Posit. The SIMD variants exhibit larger arithmetic error than their scalar counterparts, which motivates the application-level validation reported later in Section~\ref{sec:app_accuracy}. Since MSE penalizes large-magnitude deviations and aligns with the $\ell_2$ structure commonly used in DNN training objectives, it is used as the primary aggregate fidelity criterion, while MAE, NMED, and MRED are also reported for completeness.

\begin{table}[t]
\centering
\caption{{Error analysis of logarithmic Posit multipliers compared to accurate radix-4 Booth-based Posit multipliers.}}
\label{tab:lpos_error}
\renewcommand{\arraystretch}{1.25}
\setlength{\tabcolsep}{3pt}
\resizebox{\columnwidth}{!}{
\begin{tabular}{|c|l|c|c|c|c|c|}
\hline
\textbf{Format} & \textbf{Multiplier} & \textbf{Stage} & \textbf{MSE} & \textbf{MAE} & \shortstack{\textbf{NMED}\\\textbf{($\times 10^3$)}} & \shortstack{\textbf{MRED}\\\textbf{(\%)}} \\
\hline
\multirow{8}{*}{\begin{tabular}[c]{@{}c@{}}Scalar \\ 8-bit\end{tabular}}
& LP-2 (L-1) & 2 & 0.103 & 0.257 & 20.4 & 10.5 \\
& \textbf{LP-3 (L-2)} & \textbf{3} & \textbf{0.089} & \textbf{0.238} & \textbf{19.6} & \textbf{9.2} \\
& LP-3\_T4 (L-21) & 3 & 0.098 & 0.251 & 20.5 & 9.8 \\
& LP-3\_T5 (L-22) & 3 & 0.094 & 0.244 & 20.0 & 9.4 \\
& b2\_LP-2 (L-1b) & 2 & 0.109 & 0.266 & 21.3 & 11.4 \\
& b2\_LP-3 (L-2b) & 3 & 0.096 & 0.246 & 20.6 & 9.9 \\
& b2\_LP-3\_T4 (L-21b) & 3 & 0.104 & 0.262 & 21.4 & 10.5 \\
& b2\_LP-3\_T5 (L-22b) & 3 & 0.101 & 0.252 & 20.8 & 10.0 \\
\hline

\multirow{8}{*}{\begin{tabular}[c]{@{}c@{}}Scalar \\ 16-bit\end{tabular}}
& LP-4 (L-1) & 4 & 0.051 & 0.181 & 14.4 & 6.8 \\
& \textbf{LP-6 (L-2)} & \textbf{6} & \textbf{0.024} & \textbf{0.124} & \textbf{9.9} & \textbf{4.3} \\
& LP-6\_T8 (L-21) & 6 & 0.031 & 0.142 & 11.1 & 5.0 \\
& LP-6\_T10 (L-22) & 6 & 0.028 & 0.134 & 10.4 & 4.6 \\
& b3\_LP-4 (L-1b) & 4 & 0.057 & 0.191 & 15.3 & 7.3 \\
& b3\_LP-6 (L-2b) & 6 & 0.029 & 0.137 & 10.7 & 4.7 \\
& b3\_LP-6\_T8 (L-21b) & 6 & 0.036 & 0.150 & 12.1 & 5.5 \\
& b3\_LP-6\_T10 (L-22b) & 6 & 0.032 & 0.143 & 11.2 & 5.0 \\
\hline

\multirow{8}{*}{\begin{tabular}[c]{@{}c@{}}SIMD \\ (8b/16b)\end{tabular}}
& LP-4 (L-1) & 4 & 0.118 & 0.275 & 17.8 & 8.4 \\
& \textbf{LP-6 (L-2)} & \textbf{6} & \textbf{0.058} & \textbf{0.193} & \textbf{12.9} & \textbf{5.8} \\
& LP-6\_T8 (L-21) & 6 & 0.072 & 0.217 & 14.3 & 6.7 \\
& LP-6\_T10 (L-22) & 6 & 0.065 & 0.206 & 13.5 & 6.2 \\
& b3\_LP-4 (L-1b) & 4 & 0.127 & 0.287 & 18.9 & 9.1 \\
& b3\_LP-6 (L-2b) & 6 & 0.067 & 0.210 & 13.9 & 6.3 \\
& b3\_LP-6\_T8 (L-21b) & 6 & 0.081 & 0.226 & 15.2 & 7.1 \\
& b3\_LP-6\_T10 (L-22b) & 6 & 0.073 & 0.215 & 14.3 & 6.5 \\
\hline

\multirow{8}{*}{\begin{tabular}[c]{@{}c@{}}Scalar \\ 32-bit\end{tabular}}
& LP-8 (L-1) & 8 & 0.064 & 0.202 & 12.6 & 5.7 \\
& \textbf{LP-12 (L-2)} & \textbf{12} & \textbf{0.026} & \textbf{0.129} & \textbf{8.9} & \textbf{3.9} \\
& LP-12\_T16 (L-21) & 12 & 0.034 & 0.147 & 9.9 & 4.4 \\
& LP-12\_T20 (L-22) & 12 & 0.031 & 0.139 & 9.3 & 4.1 \\
& b5\_LP-8 (L-1b) & 8 & 0.071 & 0.213 & 13.4 & 6.1 \\
& b5\_LP-12 (L-2b) & 12 & 0.031 & 0.142 & 9.5 & 4.3 \\
& b5\_LP-12\_T16 (L-21b) & 12 & 0.039 & 0.155 & 10.6 & 4.8 \\
& b5\_LP-12\_T20 (L-22b) & 12 & 0.035 & 0.148 & 9.8 & 4.4 \\
\hline

\multirow{8}{*}{\begin{tabular}[c]{@{}c@{}}SIMD \\ (8b/16b/32b)\end{tabular}}
& LP-8 (L-1) & 8 & 0.148 & 0.307 & 18.9 & 8.8 \\
& \textbf{LP-12 (L-2)} & \textbf{12} & \textbf{0.123} & \textbf{0.280} & \textbf{17.4} & \textbf{8.1} \\
& LP-12\_T16 (L-21) & 12 & 0.134 & 0.295 & 18.3 & 8.5 \\
& LP-12\_T20 (L-22) & 12 & 0.128 & 0.287 & 17.7 & 8.2 \\
& b5\_LP-8 (L-1b) & 8 & 0.158 & 0.318 & 19.7 & 9.2 \\
& b5\_LP-12 (L-2b) & 12 & 0.129 & 0.288 & 17.9 & 8.4 \\
& b5\_LP-12\_T16 (L-21b) & 12 & 0.138 & 0.300 & 18.6 & 8.7 \\
& b5\_LP-12\_T20 (L-22b) & 12 & 0.133 & 0.292 & 18.1 & 8.4 \\
\hline

\end{tabular}
}
\end{table}

\begin{table}[t]
\centering
\renewcommand{\arraystretch}{1.25}
\caption{{Comparative FPGA resource consumption against state-of-the-art SIMD neural compute engines.}}
\label{tab:mac-util}
\setlength{\tabcolsep}{3pt}
\resizebox{\columnwidth}{!}{
\begin{tabular}{|c|l|c|c|c|c|c|}
\hline
\textbf{Format} & \textbf{Multiplier} & \textbf{LUTs} & \textbf{FFs} & \shortstack{\textbf{Delay}\\\textbf{(ns)}} & \shortstack{\textbf{Power}\\\textbf{(mW)}} & \shortstack{\textbf{EDP}\\\textbf{(aJ.s)}} \\
\hline

\multirow{9}{*}{\begin{tabular}[c]{@{}c@{}}Scalar \\ 8-bit\end{tabular}}
& Accurate (R4BM) & 517 & 175 & 2.69 & 93 & 0.67 \\
& LP-2 (L-1) & 414 & 141 & 1.9 & 64.3 & 0.24 \\
& LP-3 (L-2) & 438 & 149 & 2.01 & 70.1 & 0.29 \\
& LP-3\_T4 (L-21) & 409 & 139 & 1.87 & 63.2 & 0.23 \\
& LP-3\_T5 (L-22) & 416 & 141 & 1.89 & 64.6 & 0.24 \\
& b2\_LP-2 (L-1b) & 306 & 105 & 1.07 & 29.58 & 0.17 \\
& b2\_LP-3 (L-2b) & 322 & 110 & 1.15 & 33.4 & 0.24 \\
& \textbf{b2\_LP-3\_T4 (L-21b)} & \textbf{303} & \textbf{98} & \textbf{1.04} & \textbf{29.1} & \textbf{0.16} \\
& b2\_LP-3\_T5 (L-22b) & 310 & 112 & 1.1 & 30.4 & 0.19 \\
\hline

\multirow{9}{*}{\begin{tabular}[c]{@{}c@{}}Scalar \\ 16-bit\end{tabular}}
& Accurate (R4BM) & 1874 & 528 & 4.35 & 159 & 3 \\
& LP-4 (L-1) & 1495 & 412 & 2.77 & 102 & 0.79 \\
& LP-6 (L-2) & 1600 & 440 & 2.96 & 109.9 & 0.97 \\
& LP-6\_T8 (L-21) & 1478 & 406 & 2.73 & 100.4 & 0.75 \\
& LP-6\_T10 (L-22) & 1510 & 417 & 2.79 & 103.5 & 0.81 \\
& b3\_LP-4 (L-1b) & 784 & 208 & 1.86 & 76.4 & 0.53 \\
& b3\_LP-6 (L-2b) & 824 & 225 & 1.93 & 79.5 & 0.62 \\
& \textbf{b3\_LP-6\_T8 (L-21b)} & \textbf{752} & 217 & \textbf{1.83} & \textbf{73.2} & \textbf{0.48} \\
& b3\_LP-6\_T10 (L-22b) & 763 & \textbf{189} & 1.88 & 75.3 & 0.51 \\
\hline

\multirow{9}{*}{\begin{tabular}[c]{@{}c@{}}SIMD \\ (8b/16b)\end{tabular}}
& Accurate (R4BM) & 2486 & 801 & 5.1 & 214 & 5.6 \\
& LP-4 (L-1) & 1702 & 525 & 3.13 & 118.9 & 1.17 \\
& LP-6 (L-2) & 1810 & 558 & 3.35 & 127.8 & 1.45 \\
& LP-6\_T8 (L-21) & 1680 & 518 & 3.09 & 116.6 & 1.11 \\
& LP-6\_T10 (L-22) & 1716 & 530 & 3.16 & 120.5 & 1.2 \\
& b3\_LP-4 (L-1b) & 1182 & 389 & 1.82 & \textbf{59.6} & 0.67 \\
& b3\_LP-6 (L-2b) & 1260 & 406 & 1.97 & 67.2 & 0.86 \\
& \textbf{b3\_LP-6\_T8 (L-21b)} & \textbf{1157} & \textbf{353} & \textbf{1.75} & 60.8 & \textbf{0.62} \\
& b3\_LP-6\_T10 (L-22b) & 1209 & 392 & 1.80 & 62.9 & 0.69 \\
\hline

\multirow{9}{*}{\begin{tabular}[c]{@{}c@{}}Scalar \\ 32-bit\end{tabular}}
& Accurate (R4BM) & 4134 & 1580 & 10.6 & 402 & 45.2 \\
& LP-8 (L-1) & 3510 & 1330 & 4.4 & 227 & 4.4 \\
& LP-12 (L-2) & 3730 & 1415 & 4.95 & 242 & 5.9 \\
& LP-12\_T16 (L-21) & 3480 & 1320 & 4.35 & 224.5 & 4.25 \\
& LP-12\_T20 (L-22) & 3520 & 1335 & 4.4 & 227.5 & 4.45 \\
& \textbf{b5\_LP-8 (L-1b)} & \textbf{2420} & 925 & 2.53 & \textbf{113} & 3.62 \\
& b5\_LP-12 (L-2b) & 2598 & 992 & 2.92 & 128 & \textbf{3.45} \\
& b5\_LP-12\_T16 (L-21b) & 2458 & \textbf{898} & \textbf{2.47} & 116 & 3.53 \\
& b5\_LP-12\_T20 (L-22b) & 2475 & 987 & 2.51 & 119 & 3.74 \\
\hline

\multirow{9}{*}{\begin{tabular}[c]{@{}c@{}}SIMD \\ (8b/16b/32b)\end{tabular}}
& Accurate (R4BM) & 6163 & 1875 & 2.5 & 569 & 3.56 \\
& LP-8 (L-1) & 4390 & 1990 & 5.5 & 252 & 7.6 \\
& LP-12 (L-2) & 4810 & 1840 & 5.55 & 255.5 & 7.9 \\
& LP-12\_T16 (L-21) & 4310 & 1930 & 5.3 & 245.5 & 6.9 \\
& LP-12\_T20 (L-22) & 4470 & 2020 & 5.7 & 260 & 8.5 \\
& \textbf{b5\_LP-8 (L-1b)} & \textbf{3028} & 1396 & 3.16 & \textbf{126.8} & 4.22 \\
& b5\_LP-12 (L-2b) & 3349 & 1286 & 3.28 & 135.7 & 4.86 \\
& \textbf{b5\_LP-12\_T16 (L-21b)} & \textbf{3020} & 1318 & \textbf{3.04} & 128.1 & \textbf{3.94} \\
& b5\_LP-12\_T20 (L-22b) & 3142 & 1494 & 3.22 & 134.2 & 4.63 \\
\hline

\textbf{TCAS-II'24\cite{RPE}} & SIMD INT4/FP8/16/32 & 8054 & 1718 & 4.62 & 296 & 6.4\\ \hline
\textbf{TVLSI'22\cite{MPE}} & SIMD FP16/32/64 & 8065 & 1072 & 5.56 & 376 & 11.6\\ \hline
\textbf{TCAS-II'22\cite{UVMAC}} & Posit-FP8/16/32 & 5972 & 1634 & 3.74 & 499 & 7\\ \hline

\end{tabular}
}
\end{table}

\subsection{FPGA Evaluation}

Table~\ref{tab:mac-util} compares the proposed scalar and SIMD design variants against exact radix-4 Booth-based Posit multipliers and prior SIMD NCE implementations. Bold entries mark the best or text-discussed proposed values within each precision group. Relative to the exact Posit baseline, the best EULER-ADAS configurations achieve up to $3\times$, $6\times$, and $10\times$ lower energy-delay product (EDP) for the 8-bit, 16-bit, and 32-bit scalar cases, respectively.

The FPGA results isolate the effect of the three design parameters introduced in Section~\ref{sec:framework}. Increasing the logarithmic multiplier stage count improves arithmetic accuracy but increases LUT utilization, critical-path delay, and power. Operand truncation partially offsets this cost by reducing mantissa datapath width, with only a modest increase in arithmetic error. Bounding the regime field has the largest impact on the surrounding Posit datapath, reducing the cost of operand decoding and result encoding rather than only the multiplier. This confirms that encode/decode simplification is an important source of the overall FPGA savings.

The SIMD variants of EULER-ADAS require fewer LUTs than the TCAS-II'24~\cite{RPE}, TVLSI'22~\cite{MPE}, and TCAS-II'22~\cite{UVMAC} designs, while also reducing power relative to these prior SIMD engines. These gains arise from applying approximation and bounded-regime simplification across multiple stages of the Posit datapath, rather than only replacing the mantissa multiplier. The resulting FPGA implementation provides a favorable area-power trade-off for precision-reconfigurable Posit NCEs, with the application-level impact of the introduced arithmetic error evaluated in Section~\ref{sec:app_accuracy}.

\subsection{ASIC Results}

The RTL was synthesized and placed-and-routed using a Cadence RTL-to-GDS flow, with Genus for logic synthesis and Innovus for place-and-route. All ASIC results target a 28-nm CMOS HPC+ process at the TT corner, $25^\circ$C, and 0.9\,V. Timing, area, and power are reported after place-and-route.

\begin{figure}[!t]
    \centering
    \includegraphics[width=\columnwidth]{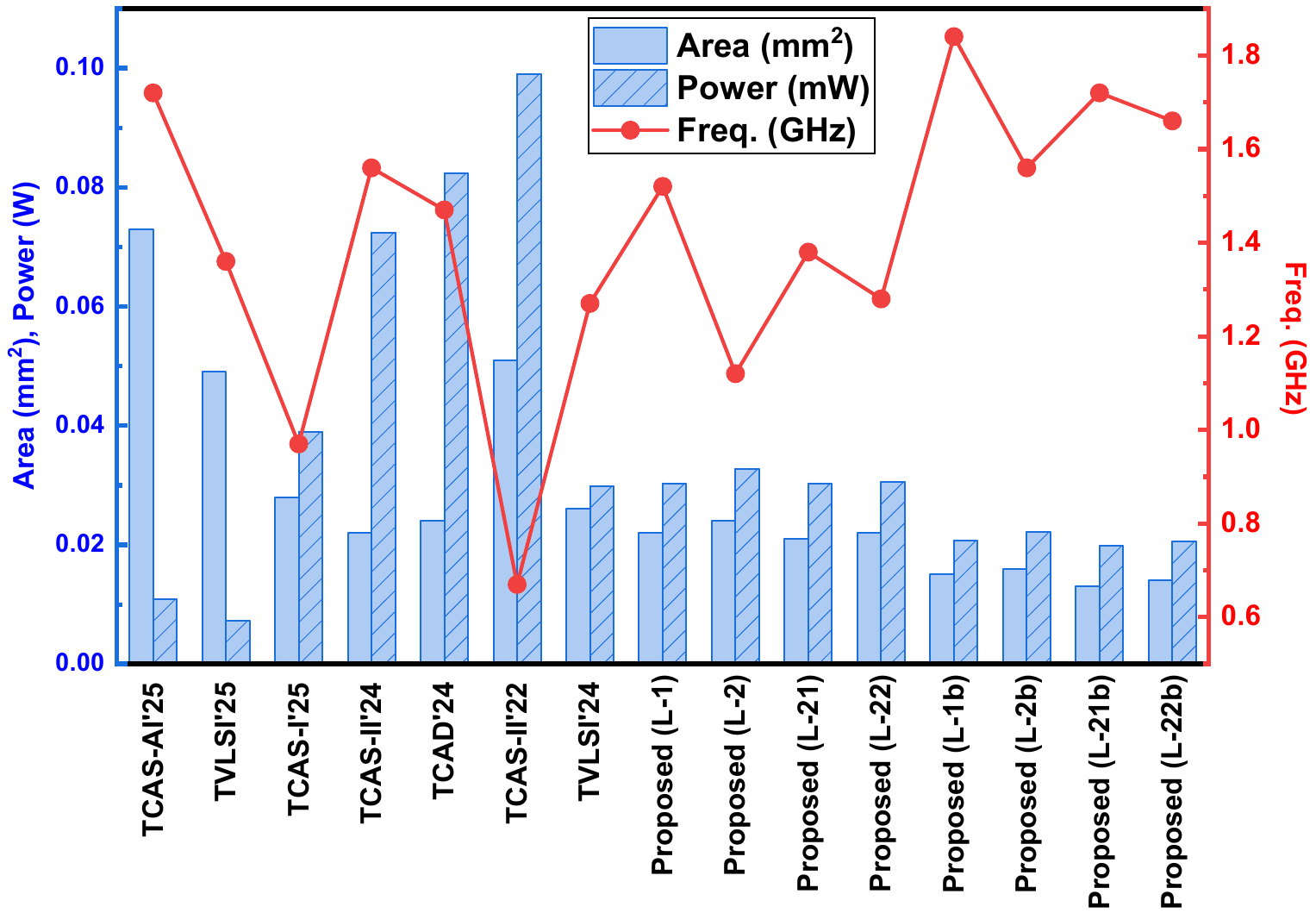}
    \caption{{Comparative 28-nm ASIC resource consumption against state-of-the-art SIMD NCEs~\cite{Flex-FMA, Flex-PE, VTA, FMA, DPDAC, UVMAC, PFP_ILM_approx, LPRE}.}}
    \label{fig:asic-mac}
\end{figure}

Fig.~\ref{fig:asic-mac} compares EULER-ADAS against state-of-the-art NCEs. Since Posit-8/16/32 can provide dynamic range comparable to FP-16/32/64 at lower bit width~\cite{RPE, Flex-FMA}, the proposed design supports a wide precision range while occupying less silicon area than the designs in~\cite{Flex-FMA, Flex-PE}. EULER-ADAS also achieves a higher operating frequency than the exact Posit design in~\cite{UVMAC} and lower power than the designs in~\cite{FMA, DPDAC}. These gains come from co-optimizing the full Posit NCE datapath: exact mantissa multiplication is replaced with stage-adaptive logarithmic multiplication, bounded-Posit representation simplifies decoding and encoding, and a shared accumulation path is preserved across all precision modes.

Table~\ref{tab:first_asic} compares error metrics and 28-nm ASIC resource consumption for prior approximate multipliers integrated into the same SIMD NCE context. Bold entries highlight the best prior reference points and the proposed values emphasized in the discussion. The proposed bounded variants (L-1b through L-22b) occupy 0.013-0.016\,mm\textsuperscript{2} and consume 19.8-22.1\,mW, reducing area and power by up to 75\% and 80\%, respectively, relative to the exact baseline. Compared with the most compact prior design in the table, ACMLC at 0.026\,mm\textsuperscript{2} and 38.9\,mW, the bounded EULER-ADAS variants reduce both area and power by approximately half. The non-bounded variants occupy 0.021-0.024\,mm\textsuperscript{2} and consume 30.3-32.7\,mW while operating at up to 1.52\,GHz. Although some proposed variants exhibit higher fixed-point MAE than prior approximate multipliers, their Posit-domain MSE remains low, indicating that large-magnitude Posit errors are limited and supporting the use of MSE as the primary arithmetic fidelity metric for DNN inference.

\begin{table}[t]
\centering
\caption{{Error-ASIC resource (28-nm) trade-off analysis: proposed SIMD NCE with state-of-the-art approximations~\cite{MITCH_TRUNC_TC, ALM_SOA, DRALM_TSC'22, HLR_BM, RAD1024_TVLSI, RoBA, TOSAM, MLC, FARA, CDM, ACE, ACSAM, HPR-Mul}.}}
\label{tab:first_asic}
\renewcommand{\arraystretch}{1.25}
\setlength{\tabcolsep}{2.5pt}
\resizebox{\columnwidth}{!}{
\begin{tabular}{|l|cc|cc|ccc|}
\hline
\textbf{Design} & \multicolumn{2}{c|}{\textbf{Fixed-point}} & \multicolumn{2}{c|}{\textbf{Posit}} & \multicolumn{3}{c|}{\textbf{SIMD NCE}} \\
\cline{2-8}
 & \shortstack{\textbf{MAE}\\\textbf{(\%)}} & \shortstack{\textbf{MSE}\\\textbf{(\%)}} & \shortstack{\textbf{MAE}\\\textbf{(\%)}} & \shortstack{\textbf{MSE}\\\textbf{(\%)}} & \shortstack{\textbf{Area}\\\textbf{(mm\textsuperscript{2})}} & \shortstack{\textbf{Freq.}\\\textbf{(GHz)}} & \shortstack{\textbf{Power}\\\textbf{(mW)}} \\
\hline

Baseline (Exact) & 0 & 0 & 0.04 & 0.09 & 0.052 & 0.67 & 99 \\
MITCH\_TRUNC & 14.43 & 1.47 & 8.65 & 0.52 & 0.023 & 1.13 & \textbf{32.7} \\
ROBA & 2.92 & 6.10 & 2.04 & 2.10 & 0.032 & 0.96 & 40.8 \\
RAD1024 & 0.44 & 1.36 & 0.31 & 0.48 & 0.029 & 1.04 & 48.3 \\
ALM\_SOA & 8.06 & 4.60 & 4.31 & 1.61 & 0.028 & 1.12 & 46.9 \\
HLR\_BM & 7.20 & 3.66 & 3.10 & 1.12 & 0.032 & 0.93 & 41.5 \\

TOSAM & 5.20 & 2.40 & 2.66 & 0.96 & 0.031 & 0.99 & 43.5 \\
HPR-Mul & 2.50 & 1.60 & 1.48 & 0.64 & 0.034 & 0.89 & 49.3 \\
ACE & 6.10 & 2.80 & 3.00 & 1.08 & 0.029 & 1.03 & 42.0 \\
ACSAM & 5.60 & 2.60 & 2.76 & 0.98 & 0.028 & 1.04 & 41.7 \\
ACMLC & 7.10 & 3.30 & 3.42 & 1.22 & \textbf{0.026} & 1.08 & 38.9 \\
CDM & 4.00 & 2.00 & 2.18 & 0.79 & 0.031 & 0.96 & 45.2 \\
FARA & 3.90 & 2.10 & 2.05 & 0.80 & 0.032 & 0.95 & 46.0 \\

\hline

Proposed (L-1) & 15.10 & 1.21 & 6.00 & 0.43 & 0.022 & \textbf{1.52} & 30.3 \\
Proposed (L-2) & 11.84 & \textbf{0.99} & \textbf{5.04} & \textbf{0.35} & 0.024 & 1.12 & 32.7 \\
Proposed (L-21) & 12.70 & 1.06 & 5.42 & 0.39 & 0.021 & 1.38 & 30.3 \\
Proposed (L-22) & 12.20 & 1.01 & 5.18 & 0.37 & 0.022 & 1.28 & 30.5 \\
Proposed (L-1b) & 15.90 & 1.27 & 6.45 & 0.47 & 0.015 & \textbf{1.84} & 20.7 \\
Proposed (L-2b) & 12.60 & 1.04 & 5.35 & 0.38 & 0.016 & 1.56 & 22.1 \\
Proposed (L-21b) & 13.35 & 1.10 & 5.82 & 0.41 & \textbf{0.013} & 1.72 & \textbf{19.8} \\
Proposed (L-22b) & 12.90 & 1.08 & 5.56 & 0.39 & 0.014 & 1.66 & 20.5 \\
\hline

\hline

\end{tabular}
}
\end{table}

Table~\ref{tab:asic_perf_metrics} reports operating frequency, power, area, throughput, energy efficiency, and compute density under fully pipelined steady-state execution. Throughput scales with SIMD lane parallelism, with the highest GOPS obtained in Posit-8 mode and proportionally lower values in Posit-16 and Posit-32 modes. Among prior approximate designs, MITCH\_TRUNC provides the highest Posit-8 throughput at 45.2\,GOPS, while ACMLC provides the best area among the listed prior designs at 0.026\,mm\textsuperscript{2}. The proposed non-bounded L-1 configuration reaches 60.8\,GOPS at 1.52\,GHz and 30.3\,mW, corresponding to 2.007\,TOPS/W and 0.276\,TOPS/mm\textsuperscript{2}. The bounded variants improve this trade-off further: L-1b reaches 73.6\,GOPS at 1.84\,GHz and 20.7\,mW, while L-21b achieves the lowest power at 19.8\,mW and the highest compute density at 0.529\,TOPS/mm\textsuperscript{2}. These text-discussed values are bolded in the table, showing that the area and power reductions from bounded encode/decode, logarithmic mantissa multiplication, and shared accumulation translate into improved system-level efficiency across precision modes. An extended ASIC performance comparison of the proposed EULER variants against SIMD approximate NCEs and Posit MPEs is provided in the Supplementary Material.

\begin{table*}[!t]
\caption{{ASIC performance metrics comparison with SIMD approximate NCEs~\cite{MITCH_TRUNC_TC, ALM_SOA, DRALM_TSC'22, HLR_BM, RAD1024_TVLSI, RoBA, TOSAM, MLC, FARA, CDM, ACE, ACSAM, HPR-Mul} and MPEs.}}
\label{tab:asic_perf_metrics}
\renewcommand{\arraystretch}{1.25}
\resizebox{\textwidth}{!}{%
\begin{tabular}{|l|c|c|c|rrr|rrr|rrr|}
\hline
\multirow{2}{*}{\textbf{Design}} & \multirow{2}{*}{\textbf{Freq (GHz)}} & \multirow{2}{*}{\textbf{Power (mW)}} & \multirow{2}{*}{\textbf{Area (mm\textsuperscript{2})}} 
& \multicolumn{3}{c|}{\textbf{Throughput (GOPS)}} 
& \multicolumn{3}{c|}{\textbf{Energy Efficiency (TOPS/W)}} 
& \multicolumn{3}{c|}{\textbf{Compute Density (TOPS/mm\textsuperscript{2})}} \\
\cline{5-13}

& & & & TP\_P8 & TP\_P16 & TP\_P32 & EE\_P8 & EE\_P16 & EE\_P32 & CD\_P8 & CD\_P16 & CD\_P32 \\
\hline

\textbf{Baseline (Exact)} & 0.67 & 99 & 0.052 & 26.8 & 12.70 & 2.82 & 0.271 & 0.129 & 0.0292 & 0.052 & 0.0244 & 0.0052 \\
\hline

MITCH\_TRUNC & 1.13 & 32.68 & 0.023 & \textbf{45.2} & 21.42 & 4.76 & 1.383 & 0.659 & 0.1490 & 0.197 & 0.0931 & 0.0209 \\
ROBA & 0.96 & 40.84 & 0.032 & 38.4 & 18.20 & 4.04 & 0.940 & 0.448 & 0.1013 & 0.120 & 0.0569 & 0.0128 \\
RAD1024 & 1.04 & 48.25 & 0.029 & 41.6 & 19.72 & 4.38 & 0.862 & 0.411 & 0.0929 & 0.143 & 0.0680 & 0.0152 \\
ALM\_SOA & 1.12 & 46.89 & 0.028 & 44.8 & 21.23 & 4.72 & 0.955 & 0.455 & 0.1029 & 0.160 & 0.0758 & 0.0168 \\
HLR\_BM & 0.93 & 41.46 & 0.032 & 37.2 & 17.63 & 3.92 & 0.897 & 0.427 & 0.0967 & 0.116 & 0.0551 & 0.0124 \\
TOSAM & 0.99 & 43.5 & 0.031 & 39.6 & 18.76 & 4.17 & 0.910 & 0.431 & 0.0958 & 0.128 & 0.0605 & 0.0135 \\
ACE & 1.03 & 42.0 & 0.029 & 41.2 & 19.52 & 4.34 & 0.981 & 0.465 & 0.1034 & 0.142 & 0.0673 & 0.0150 \\
ACSAM & 1.04 & 41.7 & 0.028 & 41.6 & 19.71 & 4.38 & 0.998 & 0.473 & 0.1049 & 0.149 & 0.0704 & 0.0156 \\
ACMLC & 1.08 & 38.9 & \textbf{0.026} & 43.2 & 20.47 & 4.55 & 1.110 & 0.526 & 0.1170 & 0.166 & 0.0787 & 0.0175 \\
CDM & 0.96 & 45.2 & 0.031 & 38.4 & 18.19 & 4.05 & 0.850 & 0.402 & 0.0895 & 0.124 & 0.0587 & 0.0131 \\
FARA & 0.95 & 46.0 & 0.032 & 38.0 & 18.00 & 4.00 & 0.826 & 0.391 & 0.0870 & 0.119 & 0.0563 & 0.0125 \\

TCAS-II'24~\cite{RPE} & 1.47 & 29.3 & 0.032 & 27.9 & 13.24 & 2.94 & 1.760 & 0.838 & 0.1896 & 0.279 & 0.1324 & 0.0294 \\
TVLSI'22~\cite{MPE} & 1.43 & 43.8 & 0.046 & 57.2 & 27.11 & 6.02 & 1.306 & 0.622 & 0.1407 & 0.124 & 0.0589 & 0.0131 \\
\hline

\textbf{Proposed (L-1)}  & \textbf{1.52} & \textbf{30.3} & \textbf{0.022} & \textbf{60.8} & \textbf{28.80} & \textbf{6.40} & \textbf{2.007} & \textbf{0.950} & \textbf{0.211} & \textbf{0.276} & \textbf{0.131} & \textbf{0.029} \\
\textbf{Proposed (L-2)}  & 1.12 & 32.7 & 0.024 & 44.8 & 21.22 & 4.72 & 1.370 & 0.649 & 0.144 & 0.187 & 0.088 & 0.020 \\
\textbf{Proposed (L-21)} & 1.38 & 30.3 & 0.021 & 55.2 & 26.15 & 5.81 & 1.822 & 0.863 & 0.192 & 0.263 & 0.125 & 0.028 \\
\textbf{Proposed (L-22)} & 1.28 & 30.5 & 0.022 & 51.2 & 24.26 & 5.39 & 1.679 & 0.796 & 0.177 & 0.233 & 0.110 & 0.025 \\
\textbf{Proposed (L-1b)} & \textbf{1.84} & \textbf{20.7} & 0.015 & \textbf{73.6} & \textbf{34.87} & \textbf{7.75} & \textbf{3.556} & \textbf{1.684} & \textbf{0.374} & \textbf{0.491} & 0.232 & 0.052 \\
\textbf{Proposed (L-2b)} & 1.56 & 22.1 & 0.016 & 62.4 & 29.56 & 6.57 & 2.824 & 1.338 & 0.297 & 0.390 & 0.185 & 0.041 \\
\textbf{Proposed (L-21b)}& 1.72 & \textbf{19.8} & \textbf{0.013} & 68.8 & 32.59 & 7.24 & 3.475 & 1.646 & 0.366 & \textbf{0.529} & \textbf{0.251} & \textbf{0.056} \\
\textbf{Proposed (L-22b)}& 1.66 & 20.5 & 0.014 & 66.4 & 31.46 & 6.99 & 3.239 & 1.535 & 0.341 & 0.474 & 0.225 & 0.050 \\ \hline

\end{tabular}}
\end{table*}

The stage-wise ASIC resource distribution in Table~\ref{tab:stage-wise_mac_asic} presents results under the typical-typical (TT) corner (0.9 V, 25$^\circ$C), while results for the slow-slow (SS) and fast-fast (FF) corners are provided in the Supplementary Material. EDP is computed as $\mathrm{EDP} = P \cdot D^{2}$, where $D = 1/f$ is the critical-path delay at maximum operating frequency $f$, and is reported in units of $10^{-5}$\,fJ$\cdot$s. Bold entries highlight the minimum proposed values in the key stage-wise resource columns. The bounded-Posit input and output processing stages reduce both area and power relative to the corresponding standard-Posit variants, confirming the benefit of simplifying variable-length encode/decode logic. In the mantissa multiplication stage, the ILM reduces area by more than 50\% and power by about 70\% relative to the TCAS-II'22 exact Posit design~\cite{UVMAC}. The comparison between L-1 and L-2 also illustrates the cost of increasing logarithmic stage count, while the truncated variants show how reducing retained mantissa width can recover part of this overhead.

\begin{table*}[t]
\centering
\caption{{Stage-wise 28-nm ASIC resource distribution against state-of-the-art SIMD neural compute engines.}}
\label{tab:stage-wise_mac_asic}
\renewcommand{\arraystretch}{1.25}
\resizebox{\textwidth}{!}{
\begin{tabular}{|c|l|l|c|c|c|c|c|c|c|}
\hline
\textbf{Design} & \textbf{Precision} & \textbf{Stage} & \textbf{S0} & \textbf{S2, S3} & \textbf{S4, S5} & \textbf{S5} & \textbf{Total} & \textbf{Max. Freq.} & \textbf{EDP} \\
 &  &  & \textbf{Input Proc.} & \textbf{Mant. Mult. \& Exp. Comp.} & \textbf{Accum.} & \textbf{Output Proc.} &  & \textbf{(GHz)} & \textbf{($10^{-5}$ fJ.s)} \\
\hline
\multirow{2}{*}{TCAD'24\cite{DPDAC}}
& \multirow{2}{*}{FP-16/32/64, BF16/TF32}
& Area ($\mu$m$^2$) & 6575 & 14735 & 3058 & 6320 & 30688 & 1.47 & 3.82 \\
&  & Power (mW) & 24.5 & 20.5 & 12.0 & 25.5 & 82.5 &  &  \\
\hline

\multirow{2}{*}{TCAS-II'22\cite{UVMAC}}
& \multirow{2}{*}{Posit-8/16/32}
& Area ($\mu$m$^2$) & 8079 & 22772 & 13273 & 5855 & 49979 & 0.67 & 22.2 \\
&  & Power (mW) & 16.2 & 43.5 & 26.0 & 14.0 & 99.7 &  &  \\
\hline

\multicolumn{1}{|c|}{\multirow{16}{*}{Proposed}} & \multicolumn{1}{l|}{\multirow{2}{*}{L. Posit-8/16/32 (L-1)}} & \multicolumn{1}{l|}{Area ($\mu$m$^2$)} & \multicolumn{1}{c|}{2156} & \multicolumn{1}{c|}{11782} & \multicolumn{1}{c|}{3058} & \multicolumn{1}{c|}{5714} & \multicolumn{1}{c|}{22710} & \multicolumn{1}{c|}{\multirow{2}{*}{1.52}} & \multirow{2}{*}{1.32} \\
\multicolumn{1}{|c|}{} & \multicolumn{1}{l|}{} & \multicolumn{1}{l|}{Power (mW)} & \multicolumn{1}{c|}{1.78} & \multicolumn{1}{c|}{11.8} & \multicolumn{1}{c|}{9.2} & \multicolumn{1}{c|}{7.52} & \multicolumn{1}{c|}{30.3} & \multicolumn{1}{c|}{} &  \\ \cline{2-10} 
\multicolumn{1}{|c|}{} & \multicolumn{1}{l|}{\multirow{2}{*}{L. Posit-8/16/32 (L-2)}} & \multicolumn{1}{l|}{Area ($\mu$m$^2$)} & \multicolumn{1}{c|}{2156} & \multicolumn{1}{c|}{13185} & \multicolumn{1}{c|}{3058} & \multicolumn{1}{c|}{5714} & \multicolumn{1}{c|}{24113} & \multicolumn{1}{c|}{\multirow{2}{*}{1.12}} & \multirow{2}{*}{2.61} \\
\multicolumn{1}{|c|}{} & \multicolumn{1}{l|}{} & \multicolumn{1}{l|}{Power (mW)} & \multicolumn{1}{c|}{1.78} & \multicolumn{1}{c|}{14.2} & \multicolumn{1}{c|}{9.2} & \multicolumn{1}{c|}{7.52} & \multicolumn{1}{c|}{32.7} & \multicolumn{1}{c|}{} &  \\ \cline{2-10} 
\multicolumn{1}{|c|}{} & \multicolumn{1}{l|}{\multirow{2}{*}{L. Posit-8/16/32 (L-21)}} & \multicolumn{1}{l|}{Area ($\mu$m$^2$)} & \multicolumn{1}{c|}{2156} & \multicolumn{1}{c|}{10353} & \multicolumn{1}{c|}{2586} & \multicolumn{1}{c|}{5714} & \multicolumn{1}{c|}{20809} & \multicolumn{1}{c|}{\multirow{2}{*}{1.38}} & \multirow{2}{*}{1.59} \\
\multicolumn{1}{|c|}{} & \multicolumn{1}{l|}{} & \multicolumn{1}{l|}{Power (mW)} & \multicolumn{1}{c|}{1.78} & \multicolumn{1}{c|}{12.4} & \multicolumn{1}{c|}{8.6} & \multicolumn{1}{c|}{7.52} & \multicolumn{1}{c|}{30.3} & \multicolumn{1}{c|}{} &  \\ \cline{2-10} 
\multicolumn{1}{|c|}{} & \multicolumn{1}{l|}{\multirow{2}{*}{L. Posit-8/16/32 (L-22)}} & \multicolumn{1}{l|}{Area ($\mu$m$^2$)} & \multicolumn{1}{c|}{2156} & \multicolumn{1}{c|}{11072} & \multicolumn{1}{c|}{2586} & \multicolumn{1}{c|}{5714} & \multicolumn{1}{c|}{21528} & \multicolumn{1}{c|}{\multirow{2}{*}{1.28}} & \multirow{2}{*}{1.86} \\
\multicolumn{1}{|c|}{} & \multicolumn{1}{l|}{} & \multicolumn{1}{l|}{Power (mW)} & \multicolumn{1}{c|}{1.78} & \multicolumn{1}{c|}{13.4} & \multicolumn{1}{c|}{7.8} & \multicolumn{1}{c|}{7.52} & \multicolumn{1}{c|}{30.5} & \multicolumn{1}{c|}{} &  \\ \cline{2-10} 
\multicolumn{1}{|c|}{} & \multicolumn{1}{l|}{\multirow{2}{*}{L. Posit-8/16/32 (L-1b)}} & \multicolumn{1}{l|}{Area ($\mu$m$^2$)} & \multicolumn{1}{c|}{\textbf{990}} & \multicolumn{1}{c|}{9285} & \multicolumn{1}{c|}{2281} & \multicolumn{1}{c|}{\textbf{2892}} & \multicolumn{1}{c|}{15448} & \multicolumn{1}{c|}{\multirow{2}{*}{\textbf{1.84}}} & \multirow{2}{*}{\textbf{0.61}} \\
\multicolumn{1}{|c|}{} & \multicolumn{1}{l|}{} & \multicolumn{1}{l|}{Power (mW)} & \multicolumn{1}{c|}{\textbf{0.82}} & \multicolumn{1}{c|}{9.3} & \multicolumn{1}{c|}{6.8} & \multicolumn{1}{c|}{\textbf{3.8}} & \multicolumn{1}{c|}{20.7} & \multicolumn{1}{c|}{} &  \\ \cline{2-10} 
\multicolumn{1}{|c|}{} & \multicolumn{1}{l|}{\multirow{2}{*}{L. Posit-8/16/32 (L-2b)}} & \multicolumn{1}{l|}{Area ($\mu$m$^2$)} & \multicolumn{1}{c|}{990} & \multicolumn{1}{c|}{9840} & \multicolumn{1}{c|}{2281} & \multicolumn{1}{c|}{2892} & \multicolumn{1}{c|}{16003} & \multicolumn{1}{c|}{\multirow{2}{*}{1.56}} & \multirow{2}{*}{0.91} \\
\multicolumn{1}{|c|}{} & \multicolumn{1}{l|}{} & \multicolumn{1}{l|}{Power (mW)} & \multicolumn{1}{c|}{0.82} & \multicolumn{1}{c|}{10.6} & \multicolumn{1}{c|}{6.8} & \multicolumn{1}{c|}{3.8} & \multicolumn{1}{c|}{22.1} & \multicolumn{1}{c|}{} &  \\ \cline{2-10} 
\multicolumn{1}{|c|}{} & \multicolumn{1}{l|}{\multirow{2}{*}{L. Posit-8/16/32 (L-21b)}} & \multicolumn{1}{l|}{Area ($\mu$m$^2$)} & \multicolumn{1}{c|}{\textbf{990}} & \multicolumn{1}{c|}{\textbf{7382}} & \multicolumn{1}{c|}{\textbf{1958}} & \multicolumn{1}{c|}{\textbf{2892}} & \multicolumn{1}{c|}{\textbf{13222}} & \multicolumn{1}{c|}{\multirow{2}{*}{1.72}} & \multirow{2}{*}{0.67} \\
\multicolumn{1}{|c|}{} & \multicolumn{1}{l|}{} & \multicolumn{1}{l|}{Power (mW)} & \multicolumn{1}{c|}{\textbf{0.82}} & \multicolumn{1}{c|}{\textbf{8.8}} & \multicolumn{1}{c|}{6.4} & \multicolumn{1}{c|}{\textbf{3.8}} & \multicolumn{1}{c|}{\textbf{19.8}} & \multicolumn{1}{c|}{} &  \\ \cline{2-10} 
\multicolumn{1}{|c|}{} & \multicolumn{1}{l|}{\multirow{2}{*}{L. Posit-8/16/32 (L-22b)}} & \multicolumn{1}{l|}{Area ($\mu$m$^2$)} & \multicolumn{1}{c|}{\textbf{990}} & \multicolumn{1}{c|}{8324} & \multicolumn{1}{c|}{\textbf{1958}} & \multicolumn{1}{c|}{\textbf{2892}} & \multicolumn{1}{c|}{14164} & \multicolumn{1}{c|}{\multirow{2}{*}{1.66}} & \multirow{2}{*}{0.74} \\
\multicolumn{1}{|c|}{} & \multicolumn{1}{l|}{} & \multicolumn{1}{l|}{Power (mW)} & \multicolumn{1}{c|}{\textbf{0.82}} & \multicolumn{1}{c|}{10.1} & \multicolumn{1}{c|}{\textbf{5.8}} & \multicolumn{1}{c|}{\textbf{3.8}} & \multicolumn{1}{c|}{20.5} & \multicolumn{1}{c|}{} &  \\ 
\hline
\end{tabular}
}
\end{table*}

\subsection{Application Accuracy and System-Level Validation}
\label{sec:app_accuracy}

Application-level impact is evaluated using bit-accurate arithmetic models of the proposed EULER-ADAS NCE variants. These models capture the bounded-regime representation, logarithmic mantissa multiplication, truncation, and SIMD operating modes used in the hardware evaluation. The resulting inference experiments quantify how the arithmetic errors reported in Table~\ref{tab:lpos_error} translate into model-level accuracy across representative workloads.

\subsubsection{Image Classification Tasks}
Table~\ref{tab:accuracy_results} presents application-level accuracy for VGG-16/CIFAR-100, ResNet-50/ImageNet, ViT-B/ImageNet, MobileNetv2/CIFAR-100, and ResNet-18/ImageNet1k, comparing the proposed EULER variants against FP32, BF16, exact Posit, and logarithmic fixed-point baselines across 8-, 16-, and 32-bit precisions.

The reference results show that exact Posit arithmetic preserves inference accuracy across the evaluated models. Exact Posit-(32,2) matches or slightly exceeds FP32, while Posit-(16,1) and Posit-(8,0) remain close to the FP32 baseline. This confirms that the underlying Posit formats provide sufficient numerical fidelity for the selected classification workloads before introducing logarithmic approximation or bounded-regime constraints.

The proposed EULER-ADAS variants follow the expected precision and approximation trends. Posit-32 configurations incur the smallest accuracy loss, with LP-12 remaining within roughly 1.1 percentage points of FP32 across the five workloads. The corresponding bounded variant adds only a small degradation, reflecting the dynamic-range reduction introduced by the regime bound. Posit-16 configurations remain within approximately 1.5 percentage points of FP32 for the best logarithmic variants, while Posit-8 exhibits a larger accuracy gap because of its limited dynamic range. The best proposed rows for each precision, together with the bounded Posit-32 counterpart discussed above, are highlighted in Table~\ref{tab:accuracy_results}. Truncated configurations fall between the lower-stage and higher-stage logarithmic variants, indicating that truncation provides a controllable area-accuracy trade-off. Across the evaluated precisions, the EULER-ADAS variants also outperform logarithmic fixed-point baselines at comparable bit widths, supporting the use of Posit arithmetic as the numerical substrate for approximate DNN inference.

\begin{table}[!t]
\caption{{Application accuracy for image classification across different precisions.}}
\label{tab:accuracy_results}
\renewcommand{\arraystretch}{1.25}
\setlength{\tabcolsep}{2.2pt}
\resizebox{\columnwidth}{!}{
\begin{tabular}{|l|l|c|c|c|c|c|}
\hline
\textbf{Design} & \textbf{Precision} & \shortstack{\textbf{VGG-16}\\\textbf{C100}} & \shortstack{\textbf{ResNet-50}\\\textbf{ImageNet}} & \shortstack{\textbf{ViT-B}\\\textbf{ImageNet}} & \shortstack{\textbf{MobileNetv2}\\\textbf{C100}} & \shortstack{\textbf{ResNet-18}\\\textbf{IN1k}} \\ \hline
\multirow{5}{*}{\textbf{Standard Ref.}} & \textbf{FP32} & 74.94 & 75.7 & 83.6 & 74.3 & 65.4 \\ \cline{2-7} 
 & \textbf{BF16} & 74.69 & 75.45 & 83.4 & 74.1 & 65.1 \\ \cline{2-7} 
 & \textbf{Posit-(8,0)} & 74.56 & 75.32 & 83.2 & 74.0 & 65.0 \\ \cline{2-7} 
 & \textbf{Posit-(16,1)} & 74.85 & 75.61 & 83.5 & 74.3 & 65.3 \\ \cline{2-7} 
 & \textbf{Posit-(32,2)} & 75.08 & 75.84 & 83.7 & 74.5 & 65.5 \\ \hline
\multirow{3}{*}{\textbf{Log-fxp\_2}} & \textbf{FxP-8} & 72.24 & 73.1 & 81.0 & 71.6 & 62.7 \\ \cline{2-7} 
 & \textbf{FxP-16} & 72.89 & 73.75 & 81.7 & 72.3 & 63.3 \\ \cline{2-7} 
 & \textbf{FxP-32} & 73.54 & 74.45 & 82.4 & 72.9 & 64.0 \\ \hline
\multirow{3}{*}{\textbf{Log-fxp\_3}} & \textbf{FxP-8} & 72.34 & 73.3 & 81.2 & 71.7 & 62.8 \\ \cline{2-7} 
 & \textbf{FxP-16} & 73.14 & 74.1 & 82.0 & 72.5 & 63.6 \\ \cline{2-7} 
 & \textbf{FxP-32} & 73.74 & 74.8 & 82.7 & 73.1 & 64.2 \\ \hline
\multirow{8}{*}{\textbf{\begin{tabular}[c]{@{}l@{}}This Work\\ (Posit-8)\end{tabular}}} & \textbf{LP-2 (L-1)} & 72.81 & 73.58 & 81.5 & 72.2 & 63.3 \\ \cline{2-7} 
 & \textbf{LP-3 (L-2)} & \textbf{73.14} & \textbf{73.9} & \textbf{81.8} & \textbf{72.5} & \textbf{63.6} \\ \cline{2-7} 
 & \textbf{LP-3\_T4 (L-21)} & 72.97 & 73.72 & 81.6 & 72.4 & 63.4 \\ \cline{2-7} 
 & \textbf{LP-3\_T5 (L-22)} & 73.05 & 73.83 & 81.7 & 72.5 & 63.5 \\ \cline{2-7} 
 & \textbf{b2\_LP-2 (L-1b)} & 72.68 & 73.41 & 81.3 & 72.1 & 63.1 \\ \cline{2-7} 
 & \textbf{b2\_LP-3 (L-2b)} & 72.95 & 73.64 & 81.5 & 72.4 & 63.4 \\ \cline{2-7} 
 & \textbf{b2\_LP-3\_T4 (L-21b)} & 72.79 & 73.53 & 81.4 & 72.2 & 63.2 \\ \cline{2-7} 
 & \textbf{b2\_LP-3\_T5 (L-22b)} & 72.87 & 73.6 & 81.5 & 72.3 & 63.3 \\ \hline
\multirow{8}{*}{\textbf{\begin{tabular}[c]{@{}l@{}}This Work\\ (Posit-16)\end{tabular}}} & \textbf{LP-4 (L-1)} & 73.26 & 74.03 & 81.9 & 72.7 & 63.7 \\ \cline{2-7} 
 & \textbf{LP-6 (L-2)} & \textbf{73.54} & \textbf{74.3} & \textbf{82.2} & \textbf{72.9} & \textbf{64.0} \\ \cline{2-7} 
 & \textbf{LP-6\_T8 (L-21)} & 73.36 & 74.17 & 82.1 & 72.8 & 63.8 \\ \cline{2-7} 
 & \textbf{LP-6\_T10 (L-22)} & 73.44 & 74.23 & 82.1 & 72.8 & 63.9 \\ \cline{2-7} 
 & \textbf{b3\_LP-4 (L-1b)} & 73.12 & 73.91 & 81.8 & 72.5 & 63.6 \\ \cline{2-7} 
 & \textbf{b3\_LP-6 (L-2b)} & 73.31 & 74.08 & 82.0 & 72.7 & 63.8 \\ \cline{2-7} 
 & \textbf{b3\_LP-6\_T8 (L-21b)} & 73.2 & 73.98 & 81.9 & 72.6 & 63.7 \\ \cline{2-7} 
 & \textbf{b3\_LP-6\_T10 (L-22b)} & 73.27 & 74.04 & 81.9 & 72.7 & 63.7 \\ \hline
\multirow{8}{*}{\textbf{\begin{tabular}[c]{@{}l@{}}This Work\\ (Posit-32)\end{tabular}}} & \textbf{LP-8 (L-1)} & 73.56 & 74.33 & 82.2 & 73.0 & 64.0 \\ \cline{2-7} 
 & \textbf{LP-12 (L-2)} & \textbf{73.84} & \textbf{74.6} & \textbf{82.5} & \textbf{73.2} & \textbf{64.3} \\ \cline{2-7} 
 & \textbf{LP-12\_T16 (L-21)} & 73.66 & 74.47 & 82.4 & 73.1 & 64.1 \\ \cline{2-7} 
 & \textbf{LP-12\_T20 (L-22)} & 73.73 & 74.52 & 82.4 & 73.1 & 64.2 \\ \cline{2-7} 
 & \textbf{b5\_LP-8 (L-1b)} & 73.42 & 74.21 & 82.1 & 72.8 & 63.9 \\ \cline{2-7} 
 & \textbf{b5\_LP-12 (L-2b)} & \textbf{73.61} & \textbf{74.36} & \textbf{82.3} & \textbf{73.0} & \textbf{64.1} \\ \cline{2-7} 
 & \textbf{b5\_LP-12\_T16 (L-21b)} & 73.52 & 74.31 & 82.2 & 72.9 & 64.0 \\ \cline{2-7} 
 & \textbf{b5\_LP-12\_T20 (L-22b)} & 73.57 & 74.34 & 82.2 & 73.0 & 64.0 \\ \hline
\end{tabular}}
\end{table}

\subsubsection{ADAS Application Scenario}
To assess the applicability of the proposed approximate logarithmic multipliers beyond image classification, we evaluate additional benchmarks grouped into seven categories covering ADAS and edge-inference workloads. Table~\ref{tab:benchmarks} summarizes the workloads used in the application-level evaluation. The benchmarks are grouped into seven categories that cover ADAS perception, driver monitoring, end-to-end vehicle control, visual-inertial odometry, augmented-reality gaze estimation, and representative edge-inference NLP tasks. The first five categories focus on autonomous-driving perception and control, including object, vehicle, and pedestrian detection; in-cabin monitoring; traffic sign recognition; road and scene perception; and PilotNet-style control from camera inputs. The remaining categories extend the evaluation to localization, gaze estimation, multilingual translation, language modeling, text-to-speech synthesis, and automatic speech recognition.


\begin{table*}[!t]
\centering
\caption{{Baseline application metrics across evaluated workloads and numerical precisions.}}
\label{tab:benchmarks}
\renewcommand{\arraystretch}{1.2}
\resizebox{\textwidth}{!}{
\begin{tabular}{|c|l|l|l|l|c|c|c|c|c|}
\hline
\textbf{Algo ID} & \textbf{Model} & \textbf{Dataset} & \textbf{Task} & \textbf{Metric} 
& \textbf{FP32} & \textbf{BF16} & \textbf{Posit-8} & \textbf{Posit-16} & \textbf{Posit-32} \\
\hline

\multicolumn{10}{|c|}{\textbf{Object / Vehicle / Pedestrian Detection}} \\
\hline
EULER-Algo\_1 & YOLO-v5 & KITTI & Object detection & mAP 
& 81.9 & 81.5 & 80.4 & 81.2 & 81.7 \\

EULER-Algo\_2 & Faster R-CNN & ETH & Vehicle detection & Accuracy 
& 92.3 & 92.0 & 91.0 & 91.8 & 92.1 \\

EULER-Algo\_3 & Custom SSD & Caltech Pedestrian & Pedestrian detection & mAP 
& 93.4 & 93.1 & 92.0 & 92.9 & 93.2 \\
 &  &  &  & F1-score 
& 0.90 & 0.89 & 0.86 & 0.89 & 0.90 \\
\hline

\multicolumn{10}{|c|}{\textbf{Driver Monitoring (In-Cabin)}} \\
\hline
EULER-Algo\_4 & HPM & ICT-3DHP & Head estimation & MAE   
& 1.84 & 1.89 & 2.05 & 1.91 & 1.86 \\

EULER-Algo\_5 & SVM + MLP & Yale B & Eye gaze / blink & Accuracy 
& 95.1 & 94.8 & 93.8 & 94.7 & 95.0 \\

EULER-Algo\_6 & Custom CNN + MLP & In-house & Driver behavior & Accuracy 
& 92.6 & 92.3 & 91.0 & 92.1 & 92.4 \\
\hline

\multicolumn{10}{|c|}{\textbf{Traffic Sign Recognition}} \\
\hline
EULER-Algo\_7 & Custom CNN & GTSRB, BTSD, STSD & Traffic classification & Accuracy 
& 83.2 & 82.9 & 81.6 & 82.6 & 83.1 \\
 &  &  &  & mAP 
& 84.7 & 84.3 & 83.1 & 84.1 & 84.5 \\
\hline

\multicolumn{10}{|c|}{\textbf{Road / Scene Perception}} \\
\hline
EULER-Algo\_8 & SAM & Cityscapes & Segmentation & mIoU 
& 96.8 & 96.5 & 95.5 & 96.3 & 96.6 \\

EULER-Algo\_9 & LaneNet & Caltech Lane & Lane detection & Accuracy 
& 91.6 & 91.3 & 90.2 & 91.1 & 91.4 \\

EULER-Algo\_10 & SDM & KITTI + Cityscapes & Depth + 3D & mAP 
& 93.1 & 92.8 & 91.6 & 92.6 & 93.0 \\
 &  &  &  & Error   
& 5.42 & 5.55 & 6.10 & 5.65 & 5.48 \\

EULER-Algo\_11 & MLP & In-house + Cityscapes & Road classification & mIoU 
& 97.1 & 96.8 & 95.8 & 96.6 & 96.9 \\
\hline

\multicolumn{10}{|c|}{\textbf{End-to-End Driving}} \\
\hline
EULER-Algo\_12 & PilotNet* & Custom dataset & Vehicle Control & MAE   
& 0.091 & 0.094 & 0.105 & 0.095 & 0.092 \\
\hline

\multicolumn{10}{|c|}{\textbf{VIO \& Eye-Gaze}} \\
\hline
EULER-Algo\_13 & UL-VIO & KITTI Odometry & VIO & Trans. Error   
& 2.18 & 2.24 & 2.45 & 2.26 & 2.21 \\
 &  &  &  & Rot. Error  
& 0.52 & 0.53 & 0.60 & 0.54 & 0.52 \\

EULER-Algo\_14 & Eye-Gaze AR & XR datasets & Gaze estimation & MSE   
& 0.0294 & 0.0305 & 0.0340 & 0.0308 & 0.0298 \\
\hline

\multicolumn{10}{|c|}{\textbf{NLP}} \\
\hline
EULER-Algo\_15 & IndicTrans2 & Multilingual & Translation & BLEU 
& 41.1 & 40.6 & 39.2 & 40.4 & 40.9 \\
 &  &  &  & NSS 
& 87.36 & 86.9 & 85.5 & 86.8 & 87.2 \\

EULER-Algo\_16 & Meta LLaMA 3.2 & Pretrained & LLM & BLEU 
& 35.2 & 34.8 & 33.5 & 34.6 & 35.1 \\

EULER-Algo\_17 & Coqui XTTS v3 & Speech & TTS & MOS 
& 4.2 & 4.15 & 3.95 & 4.12 & 4.18 \\

EULER-Algo\_18 & IndicParlerTTS & Indic speech & TTS & NSS 
& 95.36 & 94.9 & 93.5 & 94.8 & 95.1 \\
 &  &  &  & MOS 
& 4.5 & 4.45 & 4.25 & 4.42 & 4.48 \\
 &  &  &  & WER   
& 5.2 & 5.4 & 6.1 & 5.5 & 5.25 \\

EULER-Algo\_19 & NLLB-200 & Multilingual & Translation & WER   
& 2.9 & 3.0 & 3.4 & 3.1 & 2.95 \\

EULER-Algo\_20 & Whisper Large-v3 & Speech & ASR & WER   
& 2.9 & 3.0 & 3.3 & 3.1 & 2.95 \\
\hline

\end{tabular}
}
\end{table*}

Table~\ref{tab:adas_perf} presents the accuracy results across multiplier variants and numerical precisions. The same trend observed in image classification is preserved across the broader workload set. Posit-32 variants remain closest to the FP32 baseline, with the Posit-32 L-2 values highlighted in bold. Posit-16 variants generally retain near-baseline accuracy with modest degradation from logarithmic approximation and bounded-regime encoding. Operand truncation introduces a smaller penalty than reducing the logarithmic stage count, indicating that retained mantissa width can be used as a secondary tuning knob after the stage count is selected. The largest accuracy losses occur in the most aggressive Posit-8 bounded configurations, where limited dynamic range and approximation error compound. Overall, the results indicate that the proposed arithmetic configurations provide useful area and power savings while maintaining acceptable application-level behavior for the evaluated ADAS and edge-inference workloads.

\begin{table*}[!t]
\centering
\caption{{Performance evaluation for ADAS algorithms.}}
\label{tab:adas_perf}
\renewcommand{\arraystretch}{1.4}
\setlength{\tabcolsep}{2pt}
\resizebox{\textwidth}{!}{
\begin{tabular}{|l|l|cccccccc|cccccccc|cccccccc|}
\hline
\multirow{2}{*}{\textbf{Algo}} & \multirow{2}{*}{\textbf{Metric}} 
& \multicolumn{8}{c|}{\textbf{Posit-8}} 
& \multicolumn{8}{c|}{\textbf{Posit-16}} 
& \multicolumn{8}{c|}{\textbf{Posit-32}} \\
\cline{3-26}
 & 
 & L-2 & L-22 & L-21 & L-2b & L-22b & L-21b & L-1 & L-1b
 & L-2 & L-22 & L-21 & L-2b & L-22b & L-21b & L-1 & L-1b
 & L-2 & L-22 & L-21 & L-2b & L-22b & L-21b & L-1 & L-1b \\
\hline

E1 & mAP 
& 80.4 & 80.3 & 80.2 & 80.1 & 80.0 & 79.9 & 79.6 & 79.4
& 81.1 & 81.05 & 81.0 & 80.95 & 80.9 & 80.85 & 80.7 & 80.6
& \textbf{81.6} & 81.58 & 81.56 & 81.54 & 81.52 & 81.50 & 81.45 & 81.40 \\

E2 & Accuracy 
& 91.0 & 90.9 & 90.8 & 90.7 & 90.6 & 90.5 & 90.2 & 90.0
& 91.7 & 91.65 & 91.6 & 91.55 & 91.5 & 91.45 & 91.3 & 91.2
& \textbf{92.0} & 91.98 & 91.96 & 91.94 & 91.92 & 91.90 & 91.85 & 91.80 \\

E3 & mAP 
& 92.0 & 91.9 & 91.8 & 91.7 & 91.6 & 91.5 & 91.2 & 91.0
& 92.8 & 92.75 & 92.7 & 92.65 & 92.6 & 92.55 & 92.4 & 92.3
& \textbf{93.1} & 93.08 & 93.06 & 93.04 & 93.02 & 93.00 & 92.95 & 92.90 \\

E3 & F1-score 
& 0.86 & 0.85 & 0.84 & 0.83 & 0.82 & 0.81 & 0.80 & 0.78
& 0.89 & 0.885 & 0.88 & 0.875 & 0.87 & 0.865 & 0.86 & 0.855
& \textbf{0.90} & 0.898 & 0.896 & 0.894 & 0.892 & 0.890 & 0.888 & 0.885 \\

E4 & MAE 
& 2.05 & 2.10 & 2.12 & 2.15 & 2.18 & 2.20 & 2.25 & 2.30
& 1.92 & 1.94 & 1.96 & 1.98 & 2.00 & 2.02 & 2.05 & 2.08
& \textbf{1.87} & 1.88 & 1.89 & 1.90 & 1.91 & 1.92 & 1.94 & 1.96 \\

E5 & Accuracy 
& 93.8 & 93.7 & 93.6 & 93.5 & 93.4 & 93.3 & 93.0 & 92.8
& 94.6 & 94.55 & 94.5 & 94.45 & 94.4 & 94.35 & 94.2 & 94.1
& \textbf{94.9} & 94.88 & 94.86 & 94.84 & 94.82 & 94.80 & 94.75 & 94.70 \\

E6 & Accuracy 
& 91.0 & 90.9 & 90.8 & 90.7 & 90.6 & 90.5 & 90.2 & 90.0
& 92.0 & 91.95 & 91.9 & 91.85 & 91.8 & 91.75 & 91.6 & 91.5
& \textbf{92.3} & 92.28 & 92.26 & 92.24 & 92.22 & 92.20 & 92.15 & 92.10 \\

E7 & Accuracy 
& 81.6 & 81.5 & 81.4 & 81.3 & 81.2 & 81.1 & 80.8 & 80.6
& 82.5 & 82.45 & 82.4 & 82.35 & 82.3 & 82.25 & 82.1 & 82.0
& \textbf{83.0} & 82.98 & 82.96 & 82.94 & 82.92 & 82.90 & 82.85 & 82.80 \\

E7 & mAP 
& 83.1 & 83.0 & 82.9 & 82.8 & 82.7 & 82.6 & 82.3 & 82.1
& 84.0 & 83.95 & 83.9 & 83.85 & 83.8 & 83.75 & 83.6 & 83.5
& \textbf{84.4} & 84.38 & 84.36 & 84.34 & 84.32 & 84.30 & 84.25 & 84.20 \\

E8 & mIoU 
& 95.5 & 95.4 & 95.3 & 95.2 & 95.1 & 95.0 & 94.7 & 94.5
& 96.2 & 96.15 & 96.1 & 96.05 & 96.0 & 95.95 & 95.8 & 95.7
& \textbf{96.5} & 96.48 & 96.46 & 96.44 & 96.42 & 96.40 & 96.35 & 96.30 \\

E9 & Accuracy 
& 90.2 & 90.1 & 90.0 & 89.9 & 89.8 & 89.7 & 89.4 & 89.2
& 91.0 & 90.95 & 90.9 & 90.85 & 90.8 & 90.75 & 90.6 & 90.5
& \textbf{91.3} & 91.28 & 91.26 & 91.24 & 91.22 & 91.20 & 91.15 & 91.10 \\

E10 & mAP 
& 91.6 & 91.5 & 91.4 & 91.3 & 91.2 & 91.1 & 90.8 & 90.6
& 92.5 & 92.45 & 92.4 & 92.35 & 92.3 & 92.25 & 92.1 & 92.0
& \textbf{92.9} & 92.88 & 92.86 & 92.84 & 92.82 & 92.80 & 92.75 & 92.70 \\

E10 & Error 
& 6.10 & 6.20 & 6.25 & 6.30 & 6.35 & 6.40 & 6.50 & 6.60
& 5.70 & 5.75 & 5.80 & 5.85 & 5.90 & 5.95 & 6.00 & 6.05
& \textbf{5.50} & 5.52 & 5.54 & 5.56 & 5.58 & 5.60 & 5.65 & 5.70 \\

E11 & mIoU 
& 95.8 & 95.7 & 95.6 & 95.5 & 95.4 & 95.3 & 95.0 & 94.8
& 96.5 & 96.45 & 96.4 & 96.35 & 96.3 & 96.25 & 96.1 & 96.0
& \textbf{96.8} & 96.78 & 96.76 & 96.74 & 96.72 & 96.70 & 96.65 & 96.60 \\

E12 & MAE 
& 0.105 & 0.108 & 0.110 & 0.112 & 0.115 & 0.118 & 0.122 & 0.125
& 0.096 & 0.098 & 0.099 & 0.101 & 0.103 & 0.105 & 0.108 & 0.110
& \textbf{0.093} & 0.094 & 0.095 & 0.096 & 0.097 & 0.098 & 0.100 & 0.102 \\

E13 & Trans Err 
& 2.45 & 2.50 & 2.52 & 2.55 & 2.58 & 2.60 & 2.65 & 2.70
& 2.28 & 2.30 & 2.32 & 2.34 & 2.36 & 2.38 & 2.40 & 2.42
& \textbf{2.22} & 2.23 & 2.24 & 2.25 & 2.26 & 2.27 & 2.29 & 2.30 \\

E13 & Rot Err 
& 0.60 & 0.62 & 0.63 & 0.64 & 0.65 & 0.66 & 0.68 & 0.70
& 0.55 & 0.56 & 0.57 & 0.58 & 0.59 & 0.60 & 0.62 & 0.64
& \textbf{0.53} & 0.54 & 0.55 & 0.56 & 0.57 & 0.58 & 0.60 & 0.62 \\

E14 & MSE 
& 0.034 & 0.035 & 0.036 & 0.037 & 0.038 & 0.039 & 0.041 & 0.043
& 0.031 & 0.032 & 0.033 & 0.034 & 0.035 & 0.036 & 0.038 & 0.040
& \textbf{0.030} & 0.031 & 0.032 & 0.033 & 0.034 & 0.035 & 0.037 & 0.039 \\

E15 & BLEU 
& 39.2 & 39.0 & 38.8 & 38.6 & 38.4 & 38.2 & 37.8 & 37.5
& 40.3 & 40.1 & 39.9 & 39.7 & 39.5 & 39.3 & 39.0 & 38.8
& \textbf{40.8} & 40.7 & 40.6 & 40.5 & 40.4 & 40.3 & 40.1 & 40.0 \\

E15 & NSS 
& 85.5 & 85.3 & 85.1 & 84.9 & 84.7 & 84.5 & 84.0 & 83.5
& 86.7 & 86.5 & 86.3 & 86.1 & 85.9 & 85.7 & 85.4 & 85.1
& \textbf{87.1} & 87.0 & 86.9 & 86.8 & 86.7 & 86.6 & 86.4 & 86.3 \\

E16 & BLEU 
& 33.5 & 33.3 & 33.1 & 32.9 & 32.7 & 32.5 & 32.2 & 32.0
& 34.5 & 34.3 & 34.1 & 33.9 & 33.7 & 33.5 & 33.2 & 33.0
& \textbf{35.0} & 34.9 & 34.8 & 34.7 & 34.6 & 34.5 & 34.3 & 34.2 \\

E17 & MOS 
& 3.95 & 3.90 & 3.88 & 3.85 & 3.82 & 3.80 & 3.75 & 3.70
& 4.10 & 4.08 & 4.06 & 4.04 & 4.02 & 4.00 & 3.98 & 3.96
& \textbf{4.15} & 4.14 & 4.13 & 4.12 & 4.11 & 4.10 & 4.08 & 4.06 \\

E18 & NSS 
& 93.5 & 93.2 & 93.0 & 92.8 & 92.6 & 92.4 & 92.0 & 91.5
& 94.7 & 94.5 & 94.3 & 94.1 & 93.9 & 93.7 & 93.4 & 93.1
& \textbf{95.0} & 94.9 & 94.8 & 94.7 & 94.6 & 94.5 & 94.3 & 94.2 \\

E18 & MOS 
& 4.25 & 4.20 & 4.18 & 4.15 & 4.12 & 4.10 & 4.05 & 4.00
& 4.40 & 4.38 & 4.36 & 4.34 & 4.32 & 4.30 & 4.28 & 4.26
& \textbf{4.45} & 4.44 & 4.43 & 4.42 & 4.41 & 4.40 & 4.38 & 4.36 \\

E18 & WER 
& 6.10 & 6.20 & 6.25 & 6.30 & 6.35 & 6.40 & 6.50 & 6.60
& 5.60 & 5.65 & 5.70 & 5.75 & 5.80 & 5.85 & 5.90 & 5.95
& \textbf{5.30} & 5.32 & 5.34 & 5.36 & 5.38 & 5.40 & 5.45 & 5.50 \\

E19 & WER 
& 3.40 & 3.45 & 3.48 & 3.50 & 3.52 & 3.55 & 3.60 & 3.65
& 3.10 & 3.12 & 3.14 & 3.16 & 3.18 & 3.20 & 3.22 & 3.25
& \textbf{3.00} & 3.02 & 3.03 & 3.04 & 3.05 & 3.06 & 3.08 & 3.10 \\

E20 & WER 
& 3.30 & 3.35 & 3.38 & 3.40 & 3.42 & 3.45 & 3.50 & 3.55
& 3.10 & 3.12 & 3.14 & 3.16 & 3.18 & 3.20 & 3.22 & 3.25
& \textbf{3.00} & 3.02 & 3.03 & 3.04 & 3.05 & 3.06 & 3.08 & 3.10 \\

\hline
\end{tabular}
}
\end{table*}

\subsection{FPGA Prototype}

\begin{table}[!t]
\centering
\caption{Latency and power comparison for Tiny-YOLOv3/PASCAL-VOC (5.6 GOPS required per frame) across FPGA/edge platforms.}
\renewcommand{\arraystretch}{1.15}
\setlength{\tabcolsep}{3pt}
\resizebox{\columnwidth}{!}{
\begin{tabular}{|l|l|c|c|c|}
\hline
\shortstack{\textbf{Platform}\\\textbf{/ Design}} & \shortstack{\textbf{Board}\\\textbf{/ Device}} & \shortstack{\textbf{Latency}\\\textbf{(ms)}} & \shortstack{\textbf{Power}\\\textbf{(W)}} & \shortstack{\textbf{Energy}\\\textbf{(mJ/frame)}} \\
\hline
\textbf{This work (L-1)} & Pynq-Z2 & 108 & 0.44 & 47.5 \\
\textbf{This work (L-2)} & Pynq-Z2 & 128 & 0.53 & 67.8 \\ 
\textbf{This work (L-21)} & Pynq-Z2 & 104 & 0.42 & 43.8 \\
\textbf{This work (L-22)} & Pynq-Z2 & 116 & 0.48 & 55.6 \\

\textbf{This work (L-1b)} & Pynq-Z2 & 82 & 0.31 & 25.4 \\
\textbf{This work (L-2b)} & Pynq-Z2 & 95 & 0.36 & 34.2 \\
\textbf{This work (L-21b)} & Pynq-Z2 & \textbf{78} & \textbf{0.29} & \textbf{22.6} \\
\textbf{This work (L-22b)} & Pynq-Z2 & 86 & 0.33 & 28.4 \\\hline

Design-A \cite{Flex-PE} & VC707 & 186 & 2.24 & 416.6 \\
Design-Edge GPU & Jetson Nano & 226 & 1.34 & 302.8 \\
Design-MCU + NPU & STM32N6 & 195 & 0.90 & 175.5 \\
Design-CPU & Raspberry Pi & 555 & 2.70 & 1498.5 \\
Design-B \cite{Retro} & VC707 & 772 & 1.54 & 1188.9 \\
Design-Arduino-A & Portenta H7 & 460 & 2.05 & 943.0 \\
Design-Arduino-B & Nicla Vision & 520 & 2.88 & 1497.6 \\
\hline
\end{tabular}
}
\label{tab:tinyyolo_latency_power}
\end{table}

For system-level validation, the EULER-ADAS engine was deployed within the TREA architecture~\cite{LPRE} on the Pynq-Z2 for Tiny-YOLOv3 object detection. Table~\ref{tab:tinyyolo_latency_power} compares the measured latency, power, and energy per frame against FPGA and edge-platform implementations reported for the same workload. The best EULER-ADAS prototype, L-21b, achieves 78\,ms latency at 0.29\,W, corresponding to 22.6\,mJ/frame. This is lower in both latency and energy than the compared FPGA designs in~\cite{Flex-PE, Retro}, despite those designs targeting the larger VC707 platform. The Pynq-Z2 prototype also compares favorably with the listed embedded CPU, GPU, MCU, and vision-board baselines, indicating that the combined B-Posit and approximate logarithmic multiplier approach is suitable for low-power object detection at the edge.

\section{Conclusion}
\label{sec:conclusion}
This paper presented EULER-ADAS, a SIMD-enabled logarithmic Posit neural compute engine that jointly optimizes operand decoding, mantissa multiplication, quire accumulation, and result encoding for energy-efficient ADAS inference. By combining bounded-regime Posit representation, stage-adaptive logarithmic multiplication, and runtime-configurable SIMD accumulation, the proposed architecture reduces datapath cost while preserving application-level accuracy across representative workloads.

Experimental results show that the proposed bounded-Posit variants achieve up to 55\% reduction in LUT utilization, more than 45\% lower power consumption, and up to $10\times$ lower EDP compared with radix-4 Booth-based Posit multipliers. At the NCE level, the proposed configurations reduce LUT count by up to 41.4\%, delay by up to 76.1\%, and power by up to 71.9\% relative to exact Posit NCEs. In 28-nm ASIC implementation, EULER-ADAS reaches up to 1.84~GHz at 20.7~mW while occupying as little as 0.015~mm\textsuperscript{2}.

Across VGG-16, ResNet-50, ViT-B, MobileNetv2, and ResNet-18 workloads, the evaluated Posit-16 and Posit-32 configurations remain close to FP32 accuracy, confirming that the introduced arithmetic approximation has limited impact on application-level performance. The bounded-Posit variants also reduce sensitivity to catastrophic regime-field faults while lowering encode/decode complexity. System-level validation using Tiny-YOLOv3 on the Pynq-Z2 platform achieves 78~ms latency, 0.29~W power, and 22.6~mJ/frame for the best configuration, demonstrating the practical deployment potential of EULER-ADAS for low-power edge ADAS workloads.

Future work will focus on post-silicon validation, sparsity-aware and layer-adaptive runtime precision SIMD scheduling, and evaluation with a complete software stack targeting end-to-end autonomous perception pipelines for ADAS SoCs.

\bibliographystyle{ieeetr}
\bibliography{xref_bib}

\newpage

\begin{IEEEbiography}[{\includegraphics[width=1in,height=1.25in,clip,keepaspectratio]{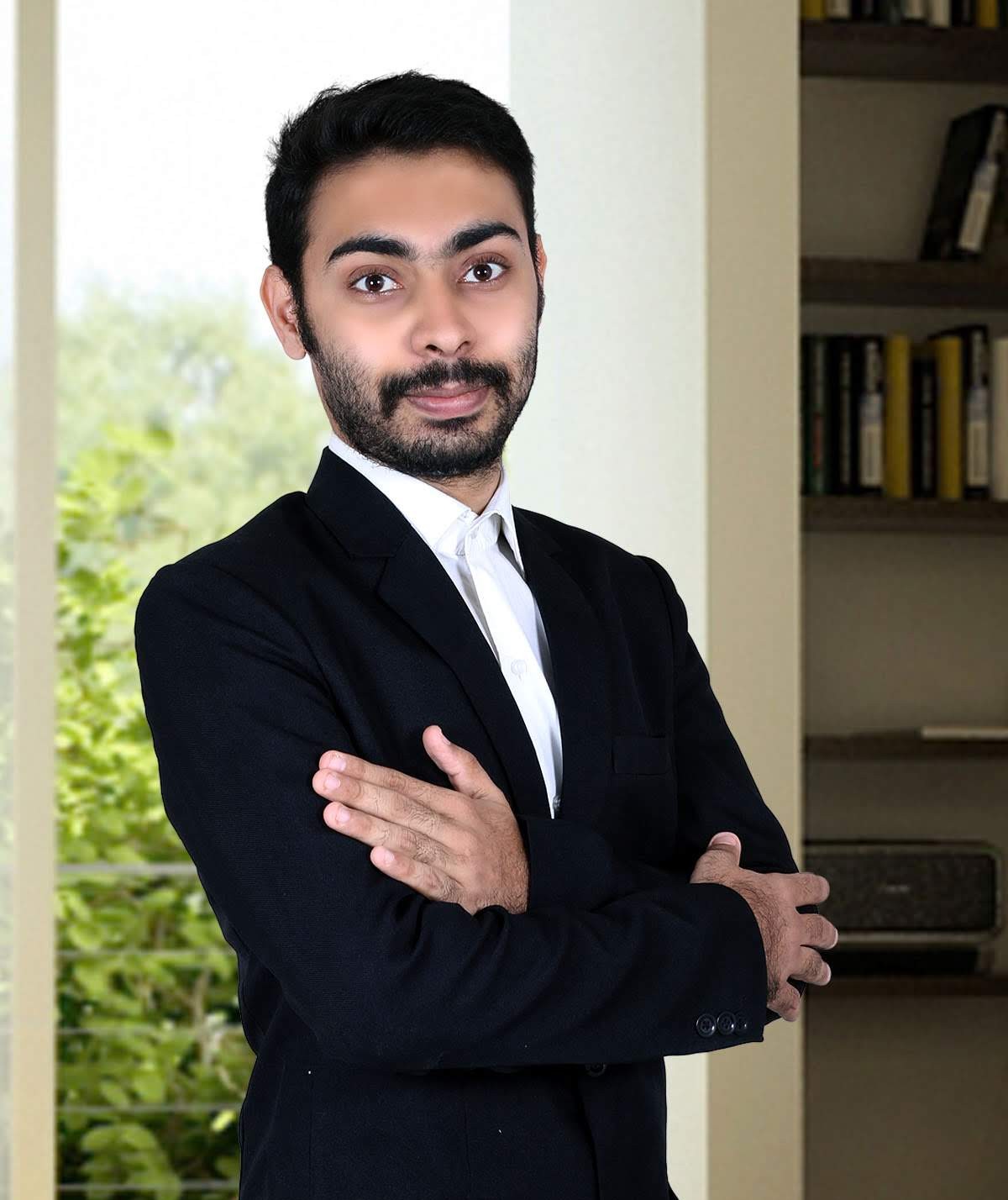}}]{Mukul Lokhande} (Member, IEEE) received the B.Tech. degree in Electronics and Communication Engineering from SGGSIE\&T, Nanded, India, in 2020. He submitted his doctoral dissertation at IIT Indore, India, in December 2025, and is currently with Qualcomm Technologies, Inc., Bengaluru, India.

His research interests include hardware-software codesign for efficient edge-AI workloads, multi-precision NPU architectures, approximate arithmetic engines, and digital processing-in-memory architectures.
\end{IEEEbiography}

\begin{IEEEbiography}[{\includegraphics[width=1in,height=1.25in,clip,keepaspectratio]{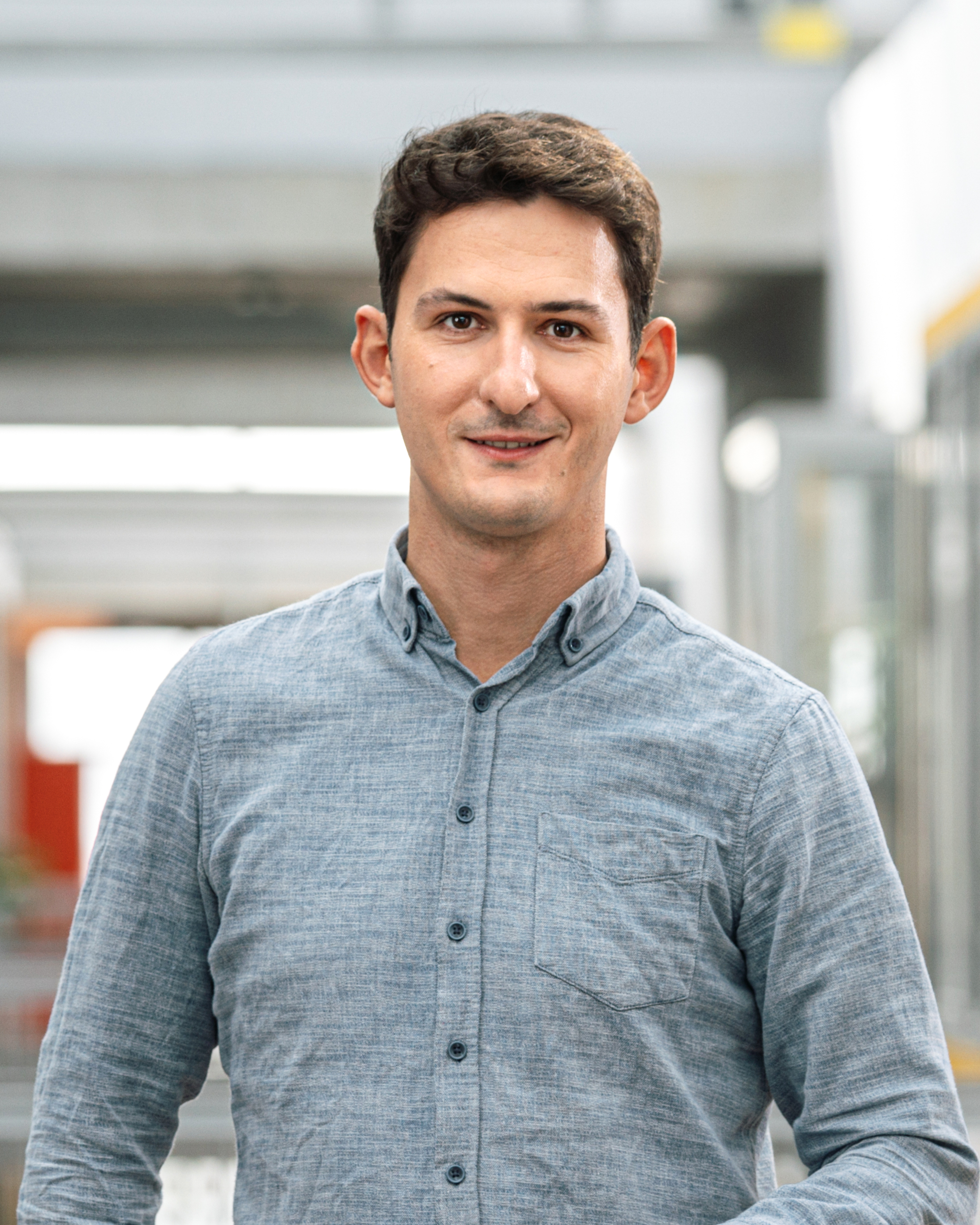}}]{Ratko Pilipovi\'{c}} received the B.Sc. and M.Sc. degrees from the Faculty of Electrical Engineering, University of Banja Luka, Bosnia and Herzegovina, in 2015 and 2017, respectively.

He received the Ph.D. degree from the Faculty of Computer and Information Science, University of Ljubljana, Slovenia, in 2021, where he is currently an Assistant Professor. His research interests include approximate computing, arithmetic circuit design, FPGA design, embedded processing, machine learning, and quantum computing.
\end{IEEEbiography}

\begin{IEEEbiography}[{\includegraphics[width=1in,height=1.25in,clip,keepaspectratio]{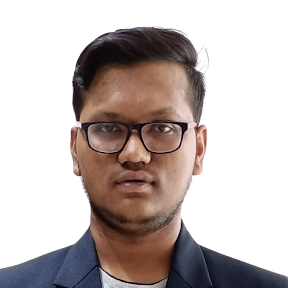}}]{Omkar Kokane} received the B.Tech. degree in Electronics and Communication Engineering from IIIT Pune, India, and the M.Tech. degree in VLSI Design and Nanoelectronics from IIT Indore, India, in 2022 and 2025, respectively.

He is currently pursuing the Ph.D. degree with the PULP Team at the University of Bologna, Italy. His research interests include matrix computation, RISC accelerator extensions, approximate computing, CORDIC-based computation, and reconfigurable design.
\end{IEEEbiography}

\begin{IEEEbiography}[{\includegraphics[width=1in,height=1.25in,clip,keepaspectratio]{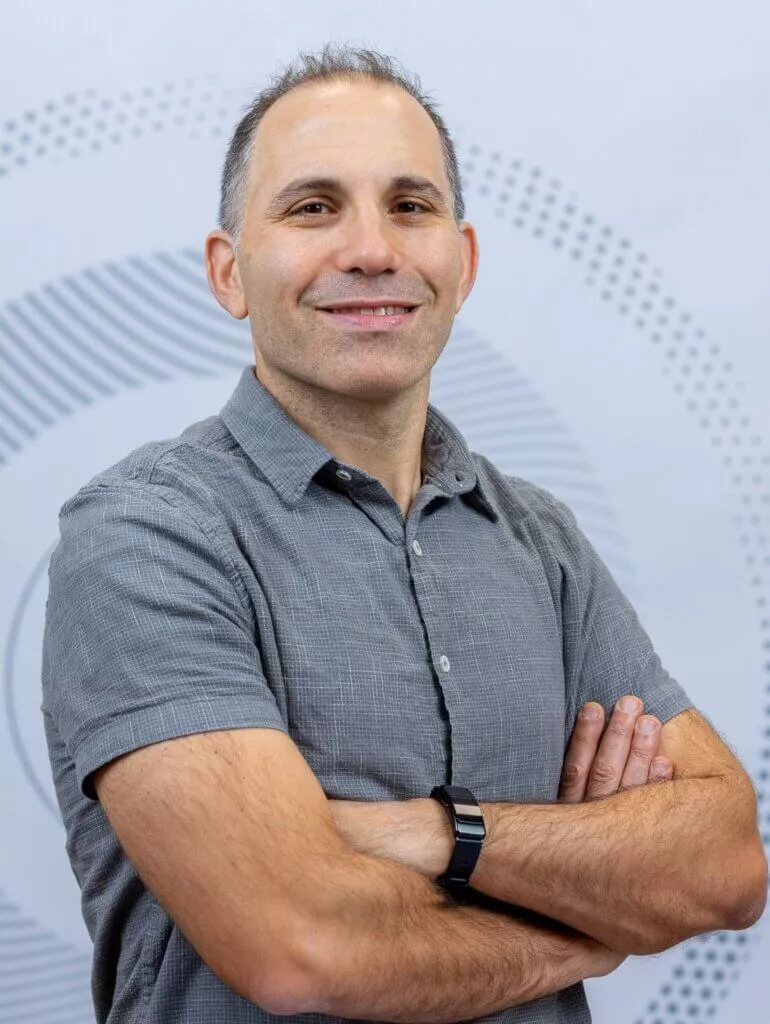}}]{Adam Teman} (Senior Member, IEEE) received the Ph.D. degree in Electrical and Computer Engineering from Ben-Gurion University of the Negev, Israel, in 2014. He is a Full Professor at Bar-Ilan University, a Co-Director of the EnICS Laboratories, and a Co-Founder of RAAAM Memory Technologies, Petach Tikva, Israel.

He has authored more than 115 publications and holds 11 patents. His research interests include embedded memories, energy-efficient circuit design, AI hardware accelerators, and physical design methodologies. He received the Krill Prize for Outstanding Young Researchers in 2020. Prof. Teman is an Associate Editor of IEEE TCAD.
\end{IEEEbiography}

\begin{IEEEbiography}[{\includegraphics[width=1in,height=1.25in,clip,keepaspectratio]{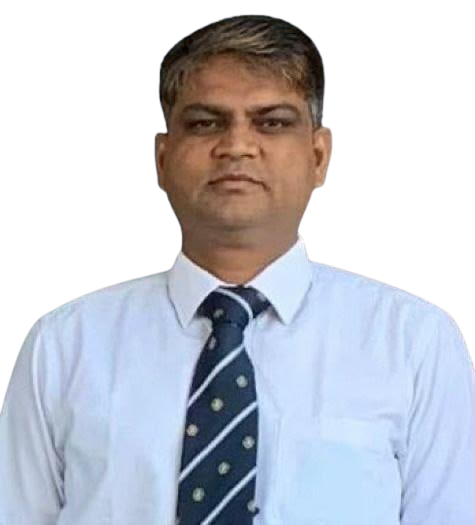}}]{Santosh Kumar Vishvakarma} (Senior Member, IEEE) received the Ph.D. degree from the Indian Institute of Technology Roorkee, India, in 2010. From 2009 to 2010, he was with the University Graduate Centre, Norway, as a Postdoctoral Fellow under a European Union project.

He is a Professor in the Department of Electrical Engineering at the Indian Institute of Technology Indore, India, and leads the Nanoscale Devices, VLSI Circuit, and System Design Laboratory. His research interests include processing-in-memory designs and resource-efficient circuits for edge-AI applications.
\end{IEEEbiography}

\end{document}